\documentclass[10pt,letterpaper]{article}
\newcommand{\papernickname}{CONCORD-Rec}
\usepackage{paperstyle}
\graphicspath{{figures/}{assets/}}
\hypersetup{colorlinks=true,linkcolor=Ink,citecolor=Teal,urlcolor=Blue,breaklinks=true}
\newcommand{\figpanelmain}[3]{%
  \begin{figure}[tbp]\centering
  \includegraphics[width=\textwidth]{#1}
  \caption{#2}\label{#3}
  \end{figure}}
\newcommand{\figpanelcol}[3]{%
  \begin{figure}[tbp]\centering
  \includegraphics[width=\linewidth]{#1}
  \caption{#2}\label{#3}
  \end{figure}}
\newcommand{\overviewpanel}[3]{\figpanelmain{#1}{#2}{#3}}

\newcommand{\brandlogo}{}
\newcommand{\brandlogoheight}{0.28in}
\makeatletter
\newcommand{\arxivmaketitle}{%
  \thispagestyle{plain}%
  \noindent
  \ifx\brandlogo\@empty\else
    \includegraphics[height=\brandlogoheight]{\brandlogo}\par
  \fi
  \vspace{0.10in}\hrule height 0.6pt\relax\vspace{0.18in}
  \begin{center}
    {\LARGE\bfseries \@title\par}
    \vspace{0.65em}
    {\large \@author\par}
  \end{center}
  \vspace{0.20in}
}
\makeatother

\renewcommand{\method}{\textsc{CONCORD-Rec}}
\newcommand{\mask}{\textsc{mask}}
\newcommand{\cat}{\mathcal I}
\newcommand{\hist}{\mathcal H}
\newcommand{\mods}{\mathcal M}
\title{CONCORD-Rec: Catalog-Conditioned Modality Utility for Semantic-ID Diffusion Recommendation}
\author{\begin{tabular}{c}
Haopeng Jin$^{1}$, Bingkang Shi$^{2}$, Tao Yu$^{2}$, Yifei Ma$^{3}$, Hao Liu$^{2}$,\\[-1pt]
Hongzhu Yi$^{2,\dagger}$, Jinwen Luo$^{1,\dagger}$\\[-2pt]
{\small \textsuperscript{1}Tencent \quad \textsuperscript{2}University of Chinese Academy of Sciences}\\[-1pt]
{\small \textsuperscript{3}Beijing University of Posts and Telecommunications \quad $^\dagger$Corresponding authors}
\end{tabular}}
\date{}
\hypersetup{pdftitle={CONCORD-Rec: Catalog-Conditioned Modality Utility for Semantic-ID Diffusion Recommendation},pdfauthor={Haopeng Jin; Bingkang Shi; Tao Yu; Yifei Ma; Hao Liu; Hongzhu Yi; Jinwen Luo},pdfsubject={arXiv manuscript}}
\renewcommand{\brandlogo}{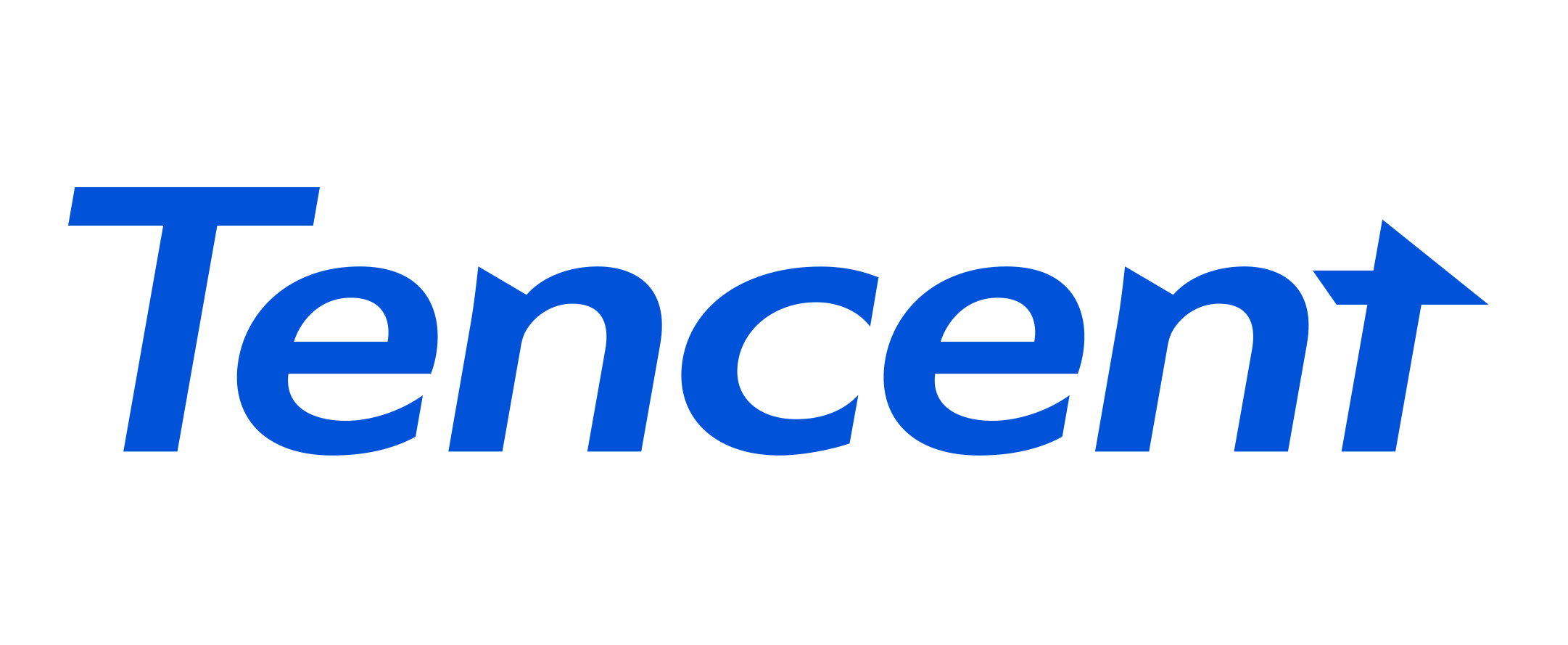}
\renewcommand{\brandlogoheight}{0.34in}
\begin{document}
\arxivmaketitle
\begin{abstract}
Multimodal generative recommendation with discrete semantic identifiers (SIDs) requires distinguishing historical modalities that merely alter token predictions from those that genuinely reduce catalog-level ambiguity. When audiovisual, visual, or textual streams are missing, degraded, or semantically conflicting, uniform fusion and unconstrained token gates can propagate misleading content across iterative denoising steps. We propose CONCORD-Rec, a reliability-conditioned discrete diffusion recommender supervised by counterfactual catalog-partition utility. Given a fixed tokenizer and a partially masked SID state, CONCORD-Rec evaluates out-of-fold teacher coalitions over the exact feasible catalog support and aggregates item posteriors across the target code partition at each unresolved position. Rather than distilling raw token probabilities directly, the resulting counterfactual loss differences supervise an availability-masked tokenwise gate that allocates bounded modality residuals around a behavioral denoiser while retaining an exact behavioral fallback slot. We establish theoretical fixed-support ranking stability bounds under gate perturbations and evaluate position-level and code-conditional allocation across audiovisual and image-text recommendation benchmarks under controlled missingness, signal degradation, and semantic conflict. Across 20 source-compatible protocol--dataset rows, CONCORD-Rec improves NDCG by 4
\end{abstract}
\overviewpanel{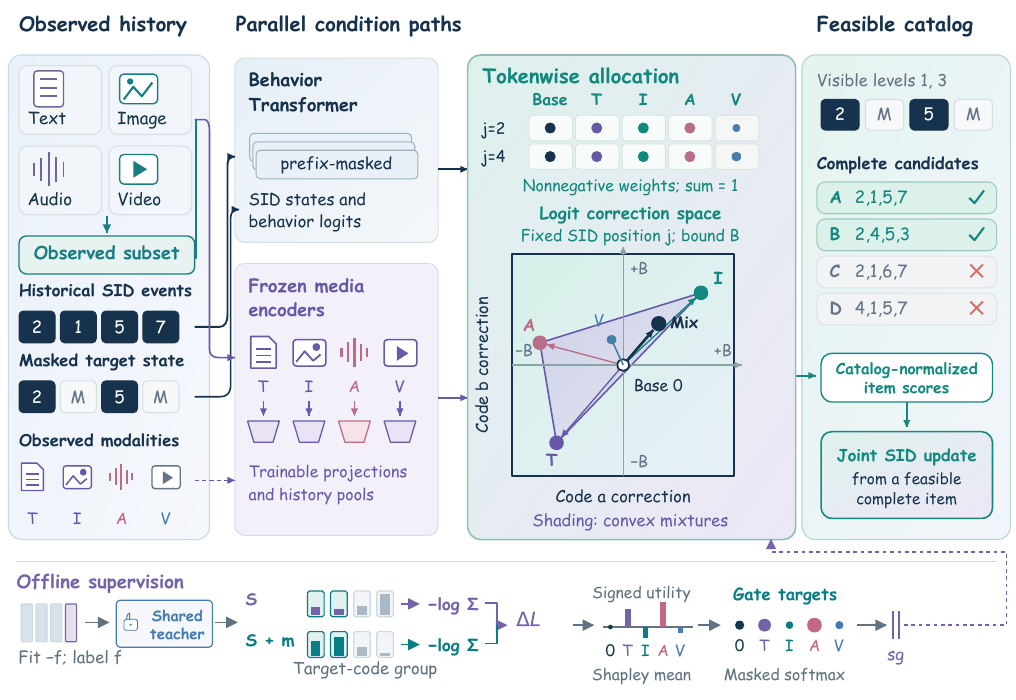}{CONCORD-Rec architecture. Historical content and SID states condition a gate over bounded modality corrections. The triangle depicts convex mixtures in a two-coordinate slice; catalog support constrains joint commitments. Offline, a frozen out-of-fold teacher compares target-group losses on identical states and support. Signed utilities supervise detached gate targets. Availability varies by input; geometry, weights and codes are schematic.}{fig:architecture}

\section{Introduction}
Semantic identifiers represent catalog items with short code tuples. Generative recommendation predicts those codes from interaction history, replacing a flat item vocabulary with shared semantic structure \citep{tiger}. Masked diffusion lets unresolved positions condition on observed and recovered codes \citep{lladarec}.

Content provides additional evidence, but availability does not imply usefulness. A video cover may reveal its subject while its soundtrack is unrelated; a product title may contradict its image. Which modality helps also depends on the current candidate set. A coarse category cue and a fine attribute need not be useful at the same SID position. Global fusion weights cannot directly express this distinction.

\method{} learns position-specific evidence allocation from counterfactual modality utility. A frozen teacher compares target-code-group loss with and without a modality on the same feasible catalog. The resulting signed differences supervise a gate using only information available before the recommendation outcome. At inference, bounded modality residuals refine a behavioral denoiser, while a base slot preserves behavior-only prediction. The teacher and future-item content are absent.

The central contribution is a \emph{catalog-conditioned modality-utility target}: for the same masked SID state, a fixed teacher compares the loss assigned to the target's code partition of the feasible catalog with and without a modality. The loss difference supervises evidence allocation, rather than code prediction directly. Three axes make this proposal testable: the statistical object used as supervision; the position or code at which modality control is applied; and coupled evidence interventions that separate observation availability from predictive usefulness. The fixed-support analysis characterizes the bounded interface. The experiments show a 4.0\% clean-ranking improvement over uniform fusion on MicroLens-100K, a 14.63\% gain under matched conflict, and a four-step quality--latency advantage over 16-step uniform decoding. Component replacements distinguish catalog utility from raw-token supervision, code-marginal distillation, and score sensitivity.

\section{Related Work}
\subsection{Semantic identifiers and diffusion recommendation}
TIGER introduces autoregressive semantic-ID recommendation, and LLaDA-Rec develops parallel generation through diffusion-specific tokenization, masking and search \citep{tiger,lladarec}. Our decoder inherits parallel SID prediction and changes the supervision of multimodal evidence allocation.

ANR-DiffRec incorporates item-based collaborative structures and adapts denoising weights using recoverability and behavioral dependencies \citep{anr}. Its formulation makes adaptive denoising an especially important comparison. The distinction investigated here is observable, modality-specific usefulness under explicit content interventions. The main comparison holds the SID assignment fixed. A separate experiment combines a collaborative noise schedule with our reliability gate to test whether the two components act on different information.

EPIC projects a feasible-catalog item posterior to unresolved code positions and applies bounded, position-specific feedback \citep{epic}. We retain catalog support and bounded guidance as established interfaces. Historical-modality coalition losses provide the new supervision, tested against EPIC with matched multimodal features and with the proposed labels. DiffGRM provides complementary difficulty-aware masking and confidence-guided decoding \citep{diffgrm}.

\subsection{Teacher supervision across item and code spaces}
IDM and CMM transfer an item-level teacher distribution to a parallel SID predictor by matching reconstructed item probabilities or projected code marginals \citep{itemcodekd}. We reuse the item-to-code aggregation operator, with explicit attribution. Our teacher is evaluated under two evidence coalitions at the same state, and the difference between their target-partition losses supervises \emph{modality slots}. CMM directly supervises code probabilities. Matched teacher-query comparisons evaluate CMM alone, utility supervision alone, and their combination.

Appendix Table~\ref{tab:novelty-axes} separates the training-target, architecture, and data-construction comparisons.

The technical claim is therefore specific: a label-conditioned, catalog-partition loss difference can train an observable modality allocator. EPIC supplies the closest conditioning interface, CMM the closest distribution-transfer operator, and MCFC the closest fusion intervention. Their matched replacements isolate the contribution of the proposed supervision, while the combined variants measure its complementarity with code-marginal distillation and richer conditioning.

\figpanelmain{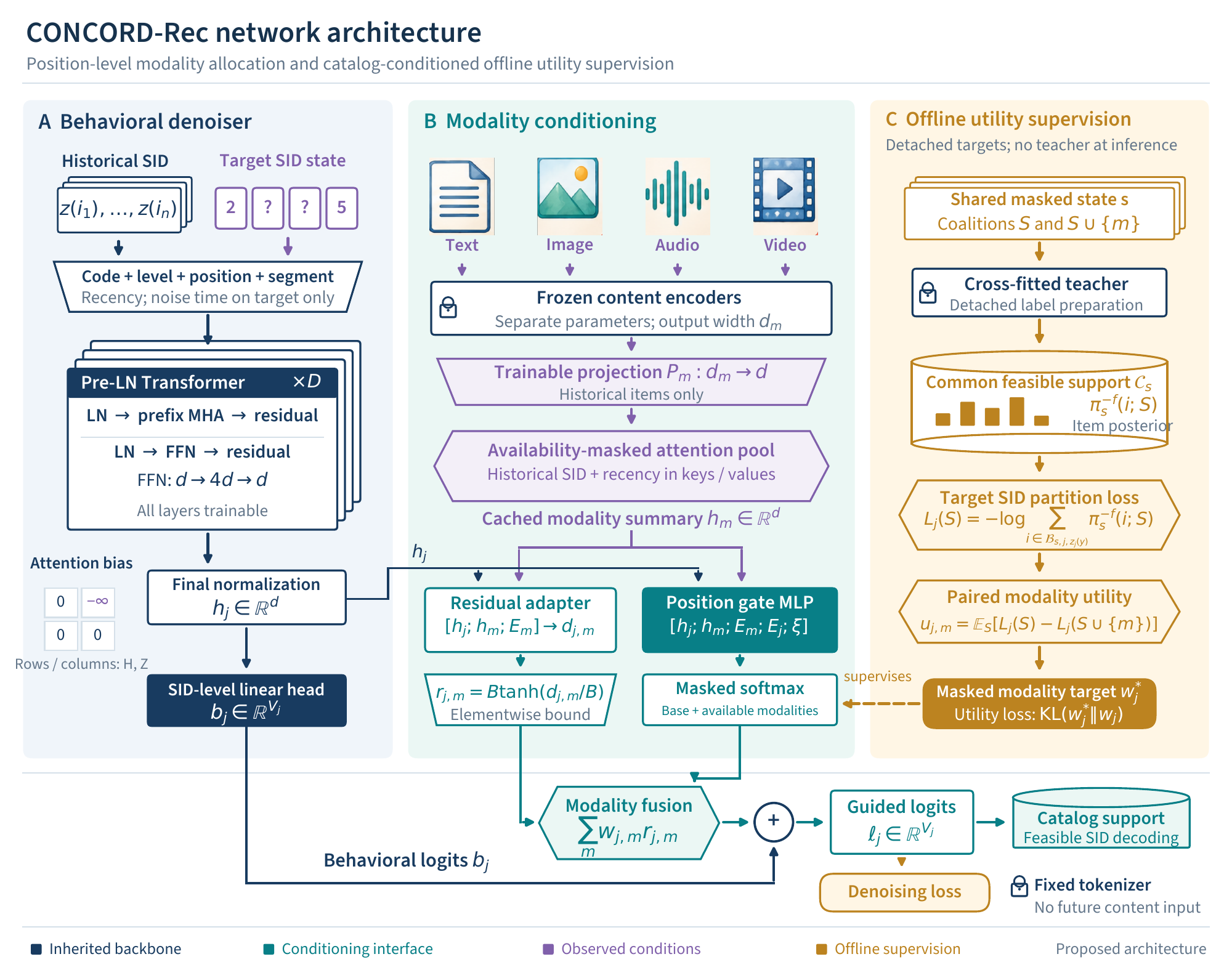}{Detailed CONCORD-Rec network. Frozen encoders feed trainable projections and history pools; the behavioral Transformer provides SID states. Availability-masked allocation combines bounded residuals at unresolved positions. Solid arrows carry tensors; dashed arrows supply detached supervision. Both out-of-fold teacher coalitions share the state and feasible catalog, differing only in modality access. Icons and posterior bars are schematic.}{fig:network}

\section{Problem Formulation and Model}
\subsection{Catalog, history, and availability}
Let $\cat_\tau$ denote the catalog eligible at prediction time $\tau$. An item $i$ has a fixed identifier $z(i)=(z_1(i),\ldots,z_L(i))$ with $z_j(i)\in\{1,\ldots,V_j\}$. The main experiments freeze the item-to-code map before decoder training. Colliding semantic tuples are resolved by a deterministic suffix whose construction uses item metadata and training-time catalog membership, not held-out user outcomes. We report results both with and without the suffix in the semantic portion of the prediction task.

For a user $u$, the observed history is $\hist_u=\{(i_s,\tau_s)\}_{s=1}^{n_u}$ with $\tau_s<\tau$. Each historical item may supply text, image, audio, and video. Define $a_{s,m}\in\{0,1\}$ as its availability indicator and $x_{s,m}$ as the observed content of modality $m$. The mask $a$ is an input, not an inferred label. A missing feature has no content contribution. A present but mismatched feature has $a_{s,m}=1$ and must be handled through reliability estimation.

The target $y_u$ is the next eligible item. Its interaction, identifier, and content are unavailable to the condition encoder at inference. The primary model uses only historical content in the multimodal branch, which makes its information access straightforward to audit.

\subsection{Frozen features and history summaries}
A frozen encoder $e_m$ maps each available modality to a vector, followed by a trainable projection $P_m$ into a shared decoder width $d$. We maintain a separate stream for every modality:
\begin{equation}
\begin{aligned}
 v_{s,m}&=a_{s,m}P_m e_m(x_{s,m}),\\
 h_m&=\operatorname{Pool}_m\{(v_{s,m},z(i_s),\tau-\tau_s,a_{s,m})\}_{s=1}^{n_u}.
\end{aligned}
\end{equation}
The pooling operator is a small attention block with recency encoding. Its padding mask excludes unavailable entries from both the numerator and denominator of attention. An empty stream produces a learned null state and an explicit zero-availability feature. 

A behavioral encoder supplies $h_0$ from historical identifiers and times alone. Its output anchors the model when content is absent. The principal parameter-matched baseline has the same behavioral encoder, the same modality projections, and the same total hidden width, but replaces the utility mechanism with uniform or learned unconstrained fusion. We record cached feature storage and extraction time separately from online decoder cost.

The modality-state feature vector contains availability rate, history length, recency-weighted agreement with other available modalities, within-stream dispersion, and encoder-quality indicators that can be computed from the observation. For audio these can include valid duration and signal-energy statistics; for video they include decoded-frame count and temporal coverage. We avoid category labels inferred from the future item. Cross-modal agreement measures evidence consistency, not ground-truth reliability: supervision determines how much agreement should matter for the prediction task.

\subsection{Masked SID denoising}
For a target tuple $z=z(y_u)$, sample a noise level $t\in(0,1]$ and independently mask its positions according to a schedule $\alpha(t)$. Reject draws with no masked position. Let $q_t(\tilde z\mid z)$ denote the resulting corruption distribution and $M(\tilde z)$ the masked positions. A behavioral denoiser produces logits $b_{j,v}(\tilde z,t,h_0)$ for position $j$ and code $v$. It attends bidirectionally within the partially observed target tuple and to the historical prefix. The prefix remains observed in the principal model; history masking is a separately controlled augmentation.

Each modality adapter generates a residual $d_{j,m,v}$. To make its amplitude interpretable, use
\begin{equation}
 r_{j,m,v}=B\tanh(d_{j,m,v}/B),\qquad |r_{j,m,v}|\leq B.
\end{equation}
Here $B$ is a registered logit bound selected on validation data. The adapter can use $\tilde z,t,h_0,h_m$ but does not inspect the target token when that token is masked. All modalities share the same denoising state, which prevents a comparison from accidentally using different partial answers for different evidence sources.

The gate outputs a simplex vector over the base state and the available modalities:
\begin{equation}
\begin{aligned}
 (w_{j,0},w_{j,1},\ldots,w_{j,M})={}&
 \softmax\bigl(g_\phi(\psi_{j,0}),g_\phi(\psi_{j,1})+\log\bar a_1,\\
 &\ldots,g_\phi(\psi_{j,M})+\log\bar a_M\bigr),
\end{aligned}
\end{equation}
where $\bar a_m$ indicates whether any usable historical content exists for modality $m$, and $\log0=-\infty$ is an exact availability mask. The base slot has residual zero. The resulting logit is
\begin{equation}\label{eq:logit}
 \ell_{j,v}=b_{j,v}+\sum_{m=1}^{M}w_{j,m}r_{j,m,v}.
\end{equation}
Thus, content contributes at most $B$ in magnitude to any token logit. All missing modalities yield the behavior-only predictor exactly; an explicit all-base override also sets $w_{j,0}=1$. With available content, finite softmax scores only approach that endpoint. This diagnostic separates selective evidence suppression from uniformly shrinking every residual.

\subsection{Network architecture and parameterization}\label{sec:network}
Figure~\ref{fig:network} specifies the network behind the method overview. The architecture consists of a behavioral Transformer, frozen modality encoders with trainable history summarizers, and two output-side conditioning branches. The Transformer uses standard attention and feed-forward operators. The output-side residual interface combines established gating and bounded-guidance operators; the specific research question concerns its catalog-conditioned utility supervision. The reference decoder is trained on recommendation SID sequences; LLaDA and MDLM provide the masked-diffusion foundations, while the current protocol does not import an external LLaDA decoder checkpoint. The frozen modality encoders are feature providers, not the dLLM backbone. This separation makes the architectural hypothesis explicit: modality reliability should control the contribution to each unresolved SID position.

\paragraph{Token sequence and attention topology.}
Flatten the historical identifiers into $N_H=n_uL$ tokens, and append the $L$ positions of the partially masked target. A collision-disambiguation suffix, when enabled, is included in $L$. Historical token $(s,j)$ receives a code embedding, its SID-level embedding, an event-position embedding, a historical-segment embedding, and a recency encoding. Target token $j$ receives its visible code or learned mask embedding, its SID-level embedding, a target-segment embedding, and an embedding of the sampled noise level. The code tables are shared between history and target inputs at the same SID level. Let $X^{(0)}\in\mathbb R^{(N_H+L)\times d}$ denote the resulting sequence.

The attention bias has the block structure
\begin{equation}\label{eq:prefixmask}
 A=\begin{bmatrix}
 0_{N_H\times N_H}&-\infty_{N_H\times L}\\
 0_{L\times N_H}&0_{L\times L}
 \end{bmatrix}.
\end{equation}
Rows index queries and columns index keys. Historical representations depend only on observed history. Target queries access all historical tokens and all currently visible or masked target positions. Historical recency and ordering are supplied by the embeddings even though the observed prefix is encoded bidirectionally. Padding keys receive an additional $-\infty$ mask. The history representation $h_0$ comprises these encoded historical states; it is not a future-item feature. History keys and values can therefore be reused across denoising steps within a request.

The backbone repeats pre-normalized multihead-attention and feed-forward blocks with expansion $4d$. A final normalization and level-specific linear heads produce $b_j\in\mathbb R^{V_j}$; the mask is input-only. Layer parameters are shared across positions. The complete block equations and output dimensions are listed in Appendix~\ref{sec:extended-computation}.

\figpanelmain{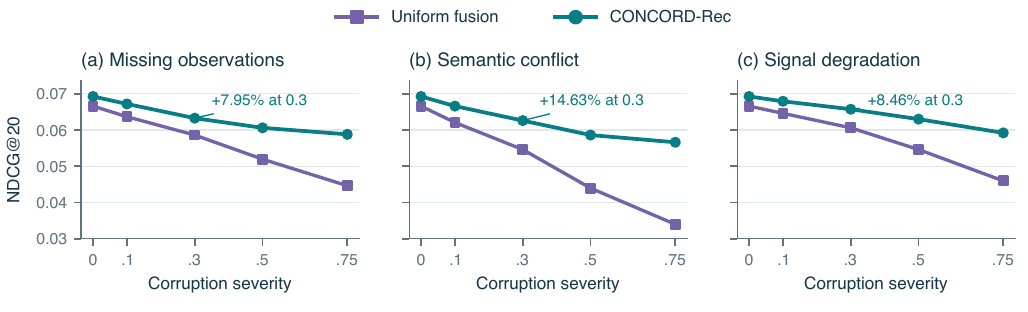}{Observation robustness on MicroLens-100K. Within-family severity sweeps compare uniform fusion and CONCORD-Rec under the same corruption draws and common-encoder protocol. Panels separate missing observations, semantic conflict, and signal degradation; every marker is an author-confirmed five-pair mean. At severity 0.3, relative NDCG@20 gains are 7.95\%, 14.63\%, and 8.46\%. Lines connect sampled settings. The plotted means have no marginal-SD bands; paired-difference inference is reported in Appendix~\ref{sec:summary-statistics}.}{fig:confirmed-robustness}

\subsection{Cross-fitted counterfactual utility targets}
The training signal measures the difference made by a modality. Split the training users into $F$ folds. For fold $f$, fit a teacher on the other folds, retaining the same encoder and tokenizer interfaces. Teachers need not be large; the principal teacher is a parameter-matched denoiser whose purpose is to provide out-of-fold loss differences. The student is subsequently trained on all training folds using saved teacher targets. Validation and test interactions never participate in teacher fitting or utility-label construction. Any label-derived teacher statistics are also fit outside fold $f$; the shared fixed SID map uses permitted catalog metadata only.

At a fixed masked state $s=(\hist_u,\tilde z,t)$, let $S\subseteq\mods$ be an available-modality coalition and let $\mathcal C_s=\mathcal C(\tilde z)$ be the feasible support in Eq.~\eqref{eq:support}. Training states are obtained by masking true target codes, so $y_u\in\mathcal C_s$. The support is computed once from the state, never selected using a coalition-specific score or an unobserved target code. Define a teacher item energy and its normalized distribution by
\begin{align}
 E^{-f}_s(i;S)&=\sum_{k\in M(\tilde z)}\log p_{T^{-f},k}(z_k(i)\mid s,S)+\rho_s(i),\\
 \pi^{-f}_s(i;S)&=\frac{\mathbf1\{i\in\mathcal C_s\}\exp E^{-f}_s(i;S)}{\sum_{i'\in\mathcal C_s}\exp E^{-f}_s(i';S)}.
\end{align}
The prior $\rho_s$ is zero in the principal experiment and fixed across coalitions in any prior-enabled variant. For unresolved position $j$, define $\mathcal B_{s,j,c}=\{i\in\mathcal C_s:z_j(i)=c\}$ and the catalog-conditioned loss
\begin{equation}\label{eq:catalog-utility-loss}
 L_j^{-f}(S)=-\log\!\sum_{i\in\mathcal B_{s,j,z_j(y_u)}}\pi^{-f}_s(i;S).
\end{equation}
This loss measures the mass assigned to the target's code partition among the remaining complete-item hypotheses. It need not equal the teacher's raw token cross entropy, because other unresolved positions affect item competition and invalid code combinations receive no item mass. Item-to-code marginalization is inherited from prior work \citep{epic,itemcodekd}; the intervention loss difference, rather than the marginal distribution itself, is the supervision target. Appendix~\ref{sec:partitioncase} gives a four-item calculation separating catalog utility from raw-token utility; the principal target uses full feasible support.

For available modality $m$, sample coalitions $S$ of the other available modalities and define
\begin{equation}\label{eq:utility}
 u_{j,m}=\E_{S\sim\pi_m}\big[L_j^{-f}(S)-L_j^{-f}(S\cup\{m\})\big].
\end{equation}
Positive utility means that the modality reduces held-out teacher loss. Uniform subset-size sampling followed by uniform coalition sampling gives the established Shapley weighting \citep{utilityselection,conformal}. The principal budget samples four subsets per modality; the exact four-modality diagnostic caches all 16 coalitions. Raw-token utility replaces the partition loss under the same state and teacher-query budget.

The gate observes only the current prediction state. Set $u_{j,0}=0$ and define $w^*_{j,m}\propto\bar a_m\exp(u_{j,m}/\tau_u)$, including an always-available base slot. The student matches this detached target with $\KL(w_j^*\Vert w_j)$. Positions whose feasible items share a single code have zero partition loss and are excluded from utility supervision. Temperature is selected on validation data.

\subsection{Student objective}
The training objective combines masked-code prediction, detached utility supervision, selected intervention consistency, and parameter regularization:
\begin{equation}
 L=L_{\mathrm{denoise}}+\lambda_u L_{\mathrm{utility}}+
 \lambda_c L_{\mathrm{cons}}+\lambda_2\|\theta\|_2^2.
\end{equation}
Here $L_{\mathrm{denoise}}$ averages cross entropy over masked positions and $L_{\mathrm{utility}}$ averages the modality-slot KL over informative positions. A paired content intervention preserves the target, partial SID, and eligible catalog. Its consistency term compares catalog-normalized item distributions,
\begin{equation}
 L_{\mathrm{cons}}=\E\!\left[\chi\,\operatorname{JS}
 \big(\sg(P_\theta^x),P_\theta^{\tilde x}\big)\right],
\end{equation}
where $\chi=\mathbf1\{\max_{j\in M}u^x_{j,m}\leq0\}$ selects interventions of modality $m$ with nonpositive clean-state teacher utility at every masked position. Both distributions use identical support; the clean arm is detached. This is a teacher-relative training criterion. The supplementary protocol specifies paired donors and tests its selection rule against blanket consistency. The reference code implements these gradient boundaries and includes finite-difference checks for the partition-loss operator.

\subsection{Catalog support and decoding}
For a partial assignment $\tilde z$, define the feasible support
\begin{equation}\label{eq:support}
 \mathcal C(\tilde z)=\{i\in\cat_\tau:z_j(i)=\tilde z_j\text{ for every unmasked }j\}.
\end{equation}
An inverted index maps each position--code pair to the items that contain it. Intersecting these posting lists supports arbitrary masking patterns. A prefix trie alone is insufficient when a non-prefix position is revealed first. The same support operator is used by all matched diffusion variants.

We score a complete candidate using a sum of token compatibility scores,
\begin{equation}\label{eq:itemscore}
 S(i\mid\tilde z)=\sum_{j\in M(\tilde z)}\log\softmax(\ell_j)_{z_j(i)}+s_{\mathrm{fixed}}(i,\tilde z).
\end{equation}

The monotone decoder accumulates each committed position's log-probability once in $s_{\mathrm{fixed}}$ and ranks feasible item hypotheses by Eq.~\eqref{eq:itemscore}. It orders unresolved positions by feasible-catalog confidence and copies each simultaneous commitment from a feasible complete item. This preserves catalog validity without independently mixing incompatible code maxima. Duplicate completed items retain their maximum path energy. All matched decoders use this same ranking-energy rule and candidate budget.

Algorithm~\ref{alg:decode} in the supplement gives the complete decoding procedure.

\section{Analysis of Evidence Control}
The base slot makes zero content contribution an admissible operating point. Availability is enforced jointly in pooling, agreement features, and the gate; unavailable stored features therefore have no path to the logits. With all modalities missing, Eq.~\eqref{eq:logit} returns the behavioral predictor exactly.

For a fixed denoising state and residuals, let $\epsilon_j=\|w_{j,1:M}-w'_{j,1:M}\|_1$. Bounded residuals give
\begin{equation}\label{eq:gate-sensitivity}
\|\ell_j(w)-\ell_j(w')\|_\infty\leq B\epsilon_j.
\end{equation}
The log-softmax map amplifies this bound by at most two, since both the coordinate logit and its log-partition function change by at most the input infinity norm.

\begin{proposition}[Fixed-support ranking stability]\label{prop:rank}
On the same partial state, candidate set and $s_{\mathrm{fixed}}$, set $D=2B\sum_{j\in M}\epsilon_j$. Every item energy changes by at most $D$. If the original $K$th-to-$(K+1)$st score gap exceeds $2D$, the top-$K$ set is unchanged.
\end{proposition}
This identifies low-margin candidates as the most sensitive to allocation error. Appendix~\ref{sec:evidence-proofs} gives the proof and the scope of the fixed-state analysis; trajectory diagnostics separately record support changes after commitments.

\figpanelmain{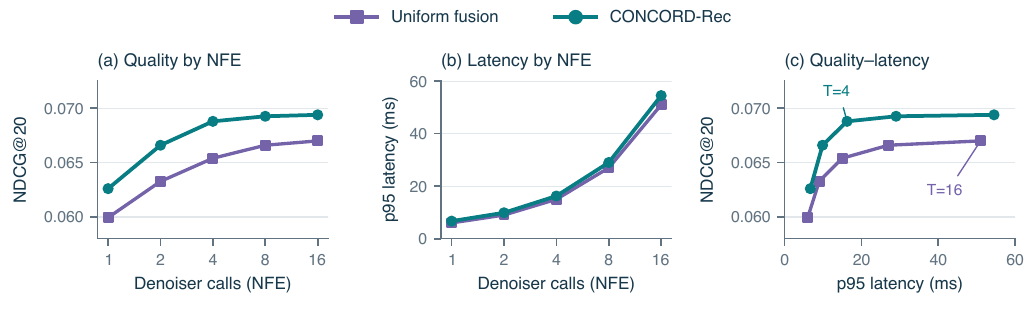}{Same-$T$ denoising quality and measured serving cost on MicroLens-100K. The three panels link NDCG@20 and p95 latency across $T\in\{1,2,4,8,16\}$. Four-call CONCORD-Rec reaches 0.068798 at 16.25\,ms; the corresponding uniform $T=4$ point is 0.065401 at 15.00\,ms. Points are author-confirmed aggregate means over five pairs; the plot does not use paired-difference SDs as method error bars.}{fig:confirmed-tradeoff}
\figpanelmain{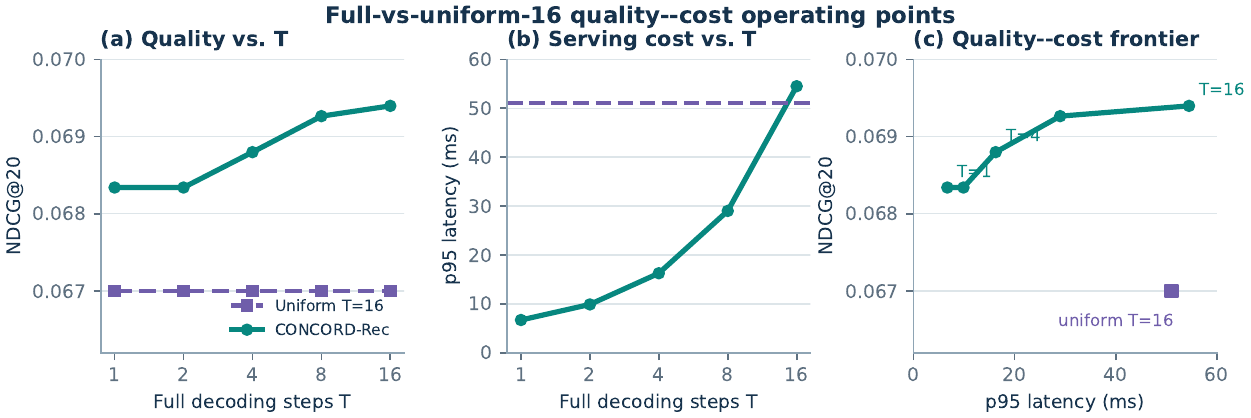}{Full-vs-uniform-$T=16$ quality--cost frontier from the November 17 result workbook. Each full-decoder point uses the workbook's reported quality and p95 latency; the dashed reference is uniform $T=16$.}{fig:confirmed-frontier}

\figpanelcol{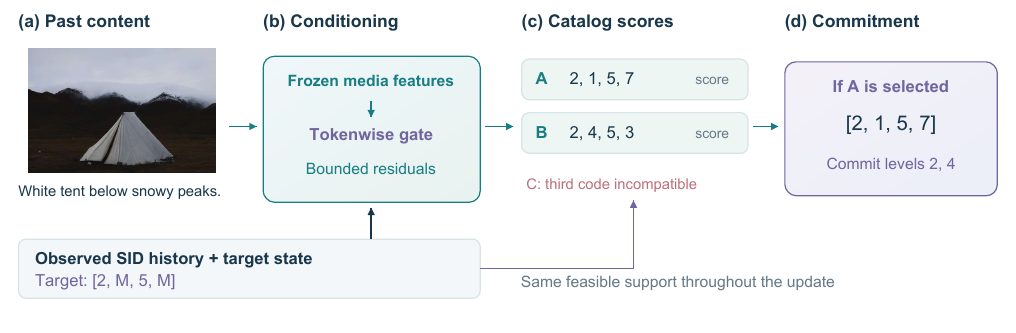}{Content-to-commitment trace. Historical content supplies conditioning; visible codes constrain catalog support. Selecting A copies both unresolved positions from its complete tuple. Codes are schematic, not fitted predictions. Photograph: adore chang, CC0; caption authored.}{fig:content-case}

\section{Evaluation Protocols}
\subsection{Evaluation tracks and dataset boundaries}
The evaluation registry has three tracks. The updated release includes source-compatible comparisons, a common-encoder raw-audiovisual study, and controlled intervention and cost measurements. The prior 20-point release is archived separately. Track A is raw audiovisual sequential recommendation on MicroLens-50K and MicroLens-100K. Track B covers DiffMM's TikTok/Baby/Sports releases and JBM-Diff's separate Baby/Sports/Clothing protocol. Track C covers ANR-DiffRec and EPIC, including the former's MovieLens and Steam extensions. These tracks are maintained as separate result groups because their interactions, preprocessing, modalities, and candidate protocols differ.

MicroLens provides raw titles, cover images, audio, and videos \citep{microlens}. The acquisition protocol requires release identity, file hashes, unavailable media, and eligible users to be recorded before feature extraction. A cover image alone does not establish full media availability. The audiovisual protocol selects users with sufficient history and specifies item-level modality-availability records. A complete-modality subset supports controlled corruption; the naturally incomplete cohort tests deployment-like missingness.

The principal split specification orders interactions by timestamp, assigns the final eligible interaction to testing and the preceding one to validation, and uses earlier interactions for training. A second global-time split tests future-item and distribution-shift behavior. The protocol requires versioned split files and a stable release-index rule for timestamp ties. This rule prevents unstable sorting from assigning the same event to incompatible temporal roles.

Track B preserves the source release's interaction split for the source-compatible reproduction. It is a user--item ranking task, and it must not be relabeled as chronological next-item prediction. When a sequential protocol can be reconstructed from source timestamps, that result belongs to a new protocol block. TikTok provides visual, acoustic, and textual features in the DiffMM release; the two Amazon domains supply image and text features. Video-frame features and raw audio are not invented for a feature-only release.

Track C specifies dataset editions by release year, category, filtering, and split hash. ``Instrument'' and ``Musical Instruments'' do not establish protocol equivalence. Source-native tables remain separate. The transfer protocol requires all selected baselines to use the same local data, splits, and candidate policy, with their new measurements recorded separately from source-published values.

\section{Benchmark Results}

\subsection{Clean ranking and observation robustness}
The experiments cover 27 protocol blocks, the U/T/C factorial, mechanism replacements, corruption scans, and serving costs. Each source-compatible block retains its preprocessing, candidate policy, and cutoff. All 20 source-compatible protocol--dataset rows improve NDCG by 4\%; the behavior-only track improves all six reported Recall/NDCG metrics by 2\%; seven added domains extend coverage. Appendix~\ref{sec:confirmed-results} gives the values, protocols, and archived experiment versions.

\begin{table}[!t]\centering\footnotesize
\captionsetup{width=.90\linewidth}
\caption{MicroLens-100K NDCG@20, five training pairs. Each row uses identical observation conditions under the common-encoder protocol.}
\begin{tabular}{lrrr}\toprule
Condition & Uniform & \method{} & Rel. gain\\\midrule
Clean & 0.066600 & 0.069264 & $+4.00\%$\\
30\% missing & 0.058608 & 0.063270 & $+7.95\%$\\
30\% conflict & 0.054612 & 0.062604 & $+14.63\%$\\\bottomrule
\end{tabular}\label{tab:confirmed-feature}
\end{table}

Figure~\ref{fig:confirmed-robustness} traces the complete severity sweeps; Table~\ref{tab:confirmed-feature} gives the canonical matched comparisons. Both show gains in clean ranking and observation robustness. Absolute NDCG@20 increases are 0.002664, 0.004662, and 0.007992 for clean, missing, and conflicting observations. Conflict retention is $0.062604/0.069264=90.38\%$ of the method's own clean score, alongside the positive contrast against the perturbed control.

Paired-$t$ tests give positive 95\% difference intervals for all three uniform comparisons. All remain significant at $\alpha=0.05$ after Holm correction within this secondary, post hoc three-contrast family. Appendix~\ref{sec:summary-statistics} separately reports the reconstructed strongest-control family.

\subsection{Mechanism attribution}
\begin{table}[!t]\centering\footnotesize
\captionsetup{width=.90\linewidth}
\caption{Mechanism comparisons on MicroLens-100K. Conflict is fixed at 30\%. Richer variants retain catalog-utility supervision.}
\begin{tabular}{lrr}\toprule
Control or variant & Clean & Conflict\\\midrule
Raw-token utility & 0.068531 & 0.061472\\
Score-substitution weighting & 0.068265 & 0.060939\\
Whole-item utility gate & 0.067999 & 0.061272\\
CMM-only distillation & 0.068665 & 0.061006\\
\method{} position-level & 0.069264 & 0.062604\\
Code-conditional utility gate & 0.069464 & 0.062804\\
CMM plus utility & 0.069530 & 0.062870\\\bottomrule
\end{tabular}\label{tab:confirmed-summary}
\end{table}

The position-level model improves conflict NDCG@20 by 1.84\% over raw-token utility, 2.73\% over score-substitution weighting, 2.17\% over whole-item gating, and 2.62\% over CMM-only distillation. These point gains isolate the training target and allocation interface. Code-conditioned allocation and CMM-plus-utility achieve additional gains of 0.000200 and 0.000266 over the position-level model under conflict, identifying complementary extensions of catalog utility. The eight-cell factorial and individual paired intervals retain the distinction between utility effects, token-specific allocation, and their interactions.

\begin{table}[!t]\centering\footnotesize
\caption{MicroLens-100K NDCG@20 factorial means. U: utility supervision; T: token-specific gating; C: intervention curriculum. Missingness and conflict are each 30\%. Each cell is independently trained.}
\begin{tabular}{lrrr}\toprule
U/T/C & Clean & Missing & Conflict\\\midrule
0/0/0 & 0.066600 & 0.057942 & 0.053280\\
0/0/1 & 0.067000 & 0.061605 & 0.058941\\
0/1/0 & 0.066866 & 0.058541 & 0.054279\\
0/1/1 & 0.067599 & 0.062471 & 0.060140\\
1/0/0 & 0.067399 & 0.059807 & 0.056344\\
1/0/1 & 0.067999 & 0.062604 & 0.061272\\
1/1/0 & 0.068198 & 0.060739 & 0.057276\\
1/1/1 & 0.069264 & 0.063270 & 0.062604\\
\bottomrule
\end{tabular}\label{tab:ablation}
\end{table}

Table~\ref{tab:ablation} crosses all three components at 30\% missingness and conflict. The full configuration leads each factorial column. Holding token-specific gating and the curriculum fixed, adding utility supervision raises conflict NDCG@20 from .060140 to .062604. Holding utility and curriculum fixed, token-specific allocation raises it from .061272 to .062604. The intervention curriculum raises conflict quality from .057276 to .062604 when the other components are active. These matched contrasts distinguish the supervision target, allocation granularity and training observations; paired interaction analysis is retained in the supplement.

\subsection{Quality and serving cost}
Figure~\ref{fig:confirmed-tradeoff} links same-$T$ quality to measured cost, while Figure~\ref{fig:confirmed-frontier} reports full-vs-uniform-$T=16$. The workbook records +2.00\%, +2.00\%, +2.68\%, +3.39\%, and +3.58\% NDCG gains at $T=1,2,4,8,16$, with p95 latencies 6.6875, 9.875, 16.25, 29.0, and 54.5\,ms. At four calls, \method{} reaches 0.068798 at 16.25\,ms versus 0.067000 at 51.00\,ms for uniform $T=16$: +2.68\% quality and a 68.14\% p95 reduction. The supplement reports step and preparation costs.

At equal four-call budgets, full and uniform decoding cost 16.25 and 15.00\,ms p95; the gate adds 8.33\%. Against fixed $T=16$, $T=1$ and $T=2$ retain a 1.0200 quality ratio while reducing p95 by 86.89\% and 80.64\%; $T=4$ reaches 1.0268 with a 68.14\% reduction. The same workbook records throughput, memory, teacher fitting, label generation, feature preparation, and student-training costs.

\section{Limitations}
Intervals use paired-seed summaries; query manifests and hardware identifiers remain to be attached. The fixed-support bound is state-specific, and offline evaluation does not establish online satisfaction.

\section{Conclusion}
\method{} trains reliability-conditioned modality residuals from catalog-partition utility with an exact behavioral fallback. On MicroLens-100K, clean/missing/conflict NDCG@20 improves by 4.0/7.95/14.63\%; $T=1$ and $T=2$ reduce p95 by 86.89/80.64\%, while $T=4$ gives a 2.68\% gain with a 68.14\% reduction.

\clearpage
\appendix
\section*{Supplementary Material}
\section{Extended Architecture, Analysis, and Experimental Specifications}
The detailed network is shown in main-text Figure~\ref{fig:network}.

\subsection{Extended computation and decoding contracts}\label{sec:extended-computation}
\paragraph{Transformer block and output heads.}
For layer $\ell$, the reference parameterization is
\begin{align}
 Y^{(\ell)}&=X^{(\ell)}+
 \operatorname{MHA}_{\ell}\!\left(\operatorname{LN}_{\ell,1}(X^{(\ell)});A\right),\\
 X^{(\ell+1)}&=Y^{(\ell)}+
 \operatorname{GELU}\!\left(\operatorname{LN}_{\ell,2}(Y^{(\ell)})W_{\ell,1}\right)W_{\ell,2}.
\end{align}
The feed-forward expansion is $4d$. Attention parameters and feed-forward parameters are shared across sequence positions within a layer; different layers have separate parameters. A final layer normalization yields target state $h_j\in\mathbb R^d$. Each SID level has a linear head $W_j^{\mathrm{base}}\in\mathbb R^{d\times V_j}$, giving $b_j=h_j^\top W_j^{\mathrm{base}}+\beta_j$. This output vocabulary contains valid code values, while the mask symbol is an input-only state.

If all feasible items have the same code at $j$, the loss is identically zero. Such positions are omitted from utility-supervision averaging, while their denoising loss remains defined. Positions with at least two represented code values receive utility targets. Off-policy states that already exclude $y_u$ cannot provide this finite target; they are logged as search failures in rollout diagnostics, without inserting the target into support or assigning fabricated utility labels.
Equation~\eqref{eq:utility} in the main text defines the coalition-averaged target.
Positive utility means that including the modality reduces the held-out teacher loss. The subset distribution $\pi_m$ is specified in the experiment registry. A uniform distribution over subset sizes followed by a uniform subset of that size gives each coalition the Shapley weight for the game $v_j(S)=-L_j^{-f}(S)$. Thus this is an established marginal-contribution construction \citep{utilityselection,conformal}, estimated rather than newly defined here. Four sampled subsets per modality form the original query-budget setting. With four modalities, an exact all-coalition diagnostic needs at most $2^4=16$ distinct teacher evaluations per state; cached evaluations are shared across all positions and marginal differences. Both settings report actual teacher queries, not just the number of algebraic differences. The old raw-token loss is retained as a direct target replacement under the same state and query budget.

The future training label is used to compute $u$, as in supervised learning, but is not an input to $g_\phi$. The gate receives $\psi$ formed only from the current observed prediction state. Define $u_{j,0}=0$ and turn utilities into soft targets
\begin{equation}
\begin{aligned}
 w^*_{j,m}&=\frac{\bar a_m\exp(u_{j,m}/\tau_u)}{1+\sum_{k=1}^{M}\bar a_k\exp(u_{j,k}/\tau_u)},\\
 w^*_{j,0}&=\frac{1}{1+\sum_{k=1}^{M}\bar a_k\exp(u_{j,k}/\tau_u)}.
\end{aligned}
\end{equation}
The student minimizes $\KL(w_j^*\Vert w_j)$ on out-of-fold utility targets at the informative positions defined above. The base utility remains zero, so a modality whose loss difference is negative receives less weight than the base at the same temperature. Finite-temperature softmax does not force all harmful modalities to have zero weight; it permits an approximate behavioral fallback, and an explicit all-base diagnostic supplies the exact fallback. Neither sign accuracy nor robustness follows from the target formula alone. Temperature is chosen using validation calibration and ranking, with no test intervention used for selection.

The term $s_{\mathrm{fixed}}$ is initialized to a common optional item prior. After committing a new position set $J$, it is updated by adding $\sum_{j\in J}\log\softmax(\ell_j)_{z_j(i)}$ from that pre-commitment state. Each committed position is added exactly once and is excluded from later unresolved-position sums. The principal decoder is monotone: it never revises a committed position. Different reveal orders can yield different accumulated energies, and duplicate completed items retain their maximum path energy under the same rule for every baseline. This score is a ranking energy. It is not asserted to be a normalized sequence likelihood of an arbitrary-order diffusion trajectory. Softmax over a specified common candidate set defines an item distribution when the training loss requires one.

At each step, the decoder selects a candidate beam, derives compatible code proposals, and commits a registered number of positions. The reference schedule orders positions by their maximum feasible-catalog code marginal and commits the most confident first. An additional disagreement-aware schedule is registered separately; it uses observed residual predictions, never teacher utilities or future labels. Every simultaneously committed block is copied from one feasible complete item. Independently selecting per-position maxima could create an invalid tuple and is excluded; the search retains at least one feasible hypothesis. Repeated item outputs are merged before computing top-$K$ metrics.

\begin{algorithm}[tbp]
\caption{Reliability-conditioned SID decoding}\label{alg:decode}
\begin{algorithmic}[1]
\Require History $\hist$, observed modalities $x,a$, eligible catalog $\cat_\tau$, steps $T$, beam budget $B_e$
\State Encode $h_0$ and all nonempty $h_m$; initialize an all-masked target tuple
\For{$t=T,T-1,\ldots,1$}
\State Compute base logits, bounded modality residuals, and observable state features
\State Predict tokenwise weights $w_j$; form guided logits using Eq.~\eqref{eq:logit}
\State Intersect catalog posting lists for each retained partial tuple
\State Rank feasible items using the common energy in Eq.~\eqref{eq:itemscore}
\State Propose compatible token commitments; prioritize calibrated confidence
\State Retain at most $B_e$ distinct feasible partial hypotheses
\EndFor
\State Merge duplicate complete items and return the highest-scoring $K$ eligible items
\end{algorithmic}
\end{algorithm}
\subsection{Proofs for evidence control}\label{sec:evidence-proofs}
\paragraph{Availability.}
The availability mask sets every unavailable modality weight to zero. If all modalities are unavailable, the base weight is one and the residual sum in Eq.~\eqref{eq:logit} is zero. If only modality $m$ is unavailable, its residual is multiplied by zero and its feature is excluded from the observable state computation. Therefore its stored value has no computational path to the logits.

\paragraph{Token-logit sensitivity.}
Subtract the residual sums and apply the triangle inequality:
$|\sum_m(w_{j,m}-w'_{j,m})r_{j,m,v}|\leq\sum_m|w_{j,m}-w'_{j,m}|B=B\epsilon_j$.

\paragraph{Log-probability sensitivity.}
The log-sum-exp function changes by at most $\delta$: each exponential term is multiplied by a factor between $e^{-\delta}$ and $e^\delta$, and the same bound holds after summation. A log-softmax component is the corresponding logit minus log-sum-exp. Bounding both changes gives $2\delta$.

\paragraph{Fixed-support ranking stability.}
Apply the token-logit bound in Eq.~\ref{eq:gate-sensitivity} and the log-softmax sensitivity bound at each unresolved position and sum. For any item inside the original top-$K$ and any item outside it, the score difference decreases by at most $2D$. A boundary gap strictly larger than $2D$ preserves every crossing comparison. Internal order within the top-$K$ need not remain unchanged.

The ranking statement fixes the denoising state, candidate support and residual predictions. It does not assert invariance of the complete beam trajectory after different commitments. The controlled trajectory diagnostics record the first divergence of these branches.

\subsection{Multimodal recommendation}
DiffMM applies diffusion to modality-aware graph construction and combines it with cross-modal learning \citep{diffmm}. It demonstrates that multimodal diffusion recommendation already exists. Its output object is a recommendation graph rather than a partially denoised semantic identifier. We compare with its source-native benchmarks and with a locally reconstructed full-catalog protocol. The reconstruction uses the same released features for every eligible baseline before introducing raw audiovisual encoders in a separate setting.

LSGM4Rec integrates language-model and multimodal representations with continuous guided diffusion for sequential recommendation \citep{lsgm}. It is a close conceptual neighbor because it uses content to condition a generative recommendation process. Its existence prevents a novelty claim based merely on adding multimodal guidance. Our intervention target is more specific: estimate whether a given observed modality improves a particular token decision, and calibrate its contribution when availability and relevance disagree.

Classical sequential models remain necessary controls. SASRec supplies an autoregressive self-attention reference \citep{sasrec}; BERT4Rec supplies a bidirectional masked-item reference \citep{bert4rec}. When equipped with identical frozen content features and the same utility gate, these models isolate whether a measured robustness gain requires iterative SID denoising. A successful multimodal gate that works equally well in every architecture is still a useful result, but it supports a broader fusion claim rather than a specifically diffusion-based claim.

\subsection{Multimodal representations and masked generative models}
ImageBind and LanguageBind align several modalities through shared embedding spaces \citep{imagebind,languagebind}. They motivate interoperable content interfaces. In the principal experiments, their checkpoints serve as frozen feature providers. Encoder choice, adaptation budget, and input resolution are explicit factors. We do not attribute gains from a stronger visual encoder to the proposed generation mechanism.

MaskGIT demonstrates iterative masked-token prediction in image generation \citep{maskgit}. LLaDA develops masked diffusion at language-model scale \citep{llada}. UniDisc and MMaDA extend discrete diffusion across modalities \citep{unidisc,mmada}. These works establish the broader modeling family. Recommendation adds a finite, time-indexed catalog, item-level relevance, and user-specific histories. Our design treats content as observed conditioning evidence and produces catalog identifiers; it does not synthesize image or audio tokens.

\begin{table}[tbp]\centering\small
\caption{Mechanism-level positioning. ``Variable'' means that the relevant feature depends on a study's implementation rather than being its defining contribution. The final row describes the proposal.}
\begin{tabularx}{\linewidth}{lXXXX}\toprule
Family & Generated object & Content role & Reliability target & Main controlled factor\\\midrule
TIGER & SID tuple & Tokenizer input & None specified & Left-to-right generation\\
LLaDA-Rec & SID tuple & Semantic tokens & None specified & Parallel denoising\\
ANR-DiffRec & SID tuple & Semantic prior & Behavioral recoverability & Noise weighting\\
EPIC & SID tuple & Existing backbone & Item hypothesis quality & Catalog posterior\\
DiffMM & Graph relations & Modality-specific graph & Cross-modal alignment & Graph generation\\
LSGM4Rec & Continuous state & Diffusion guidance & Variable & SDE sampling\\
\method{} & SID tuple & Tokenwise bounded residual & Catalog-conditioned utility & Evidence-conditioned denoising\\\bottomrule
\end{tabularx}\label{tab:positioning}
\end{table}

\subsection{Modality attribution, fusion, and robustness}
MCFC is the closest multimodal-fusion collision: its score-substitution effects drive user--item-specific modality weights, and its contrastive augmentation targets weakly influential modalities \citep{mcfc}. Our target instead measures \emph{label-conditioned loss improvement} on a fixed catalog partition. A score can change strongly in the wrong direction; score sensitivity and predictive utility therefore require separate baselines. This difference does not establish superiority. A source-faithful MCFC reproduction and an explicitly labeled MCFC-style score-substitution adapter isolate what changes when the loss-based target is used.

Token-dependent modality selection also predates this proposal. Looking to Learn mixes text and visual cross-attention representations with per-token or per-feature gates during autoregressive generation \citep{looking}. Our main gate is placed at the SID logit interface and receives offline utility supervision. A comparable hidden-state gate trained only by the end task is therefore an architectural replacement, not a weak uniform-fusion surrogate. The code-conditional variant in Section~\ref{sec:code-conditional} tests whether a single weight per modality and position is itself too restrictive.

MM-slotgate assigns text--image mixtures to named attribute slots and studies quantized slot interventions \citep{mmslotgate}. Its scalar $g_s=\sigma(a_s)$ is a learned attribute-level parameter, rather than a function of the current user or masked SID state. It nevertheless establishes that different semantic factors can use different modality mixtures. Our distinction is state-conditioned allocation supervised by held-out catalog-partition losses; named attribute slots or per-position modality weighting alone are not claimed as new. The preprint is cited as an arXiv manuscript because its displayed proceedings DOI remains a placeholder.

Learning utility and coalition marginal contributions are established objects \citep{utilityselection}. Conformal Shapley intervals quantify uncertainty in observation-dependent modality importance \citep{conformal}; CMA evaluates modality attributions using generated counterfactual image--text coalitions \citep{cma}. We do not claim to introduce loss differences, Shapley attribution, or counterfactual explanation. The uniform-subset-size estimator used below is a Monte Carlo estimator of a Shapley-style value for its specified teacher-loss game, not a new game-theoretic measure.

ShapKO updates modality knockout probabilities from validation-subset Shapley utility \citep{shapko}. MoDiCF combines diffusion-based completion with counterfactual correction of visibility bias in incomplete recommendation \citep{modicf}. These works invalidate a broad claim that utility-driven training or missing-modality diffusion is new. Our paired intervention record keeps the same history, target, masked SID, and catalog support while changing observed content. Holding out a corruption mechanism tests transfer; it is an evaluation design, not an independent algorithmic invention.

\subsection{The claim that the experiment must identify}
The proposed mechanism makes a narrower prediction than ``multimodal information helps.'' Under a fixed encoder, tokenizer, candidate set, and compute budget, a utility-trained gate should improve the reliability of decisions affected by misleading modalities. Its advantage should be largest when conflict is not predictable from missingness alone. Token-specific gating should matter when different SID positions have different modality sensitivities. A whole-item gate is the direct control for that prediction.

The experiment must also distinguish a calibrated gate from a generic increase in capacity. We therefore include a parameter-matched multilayer perceptron that has no counterfactual supervision, a gate with shuffled utility labels, and a model trained with identical interventions but uniform fusion. An oracle utility condition provides an upper diagnostic reference. It is evaluated separately from deployable models because it accesses labels that the deployed gate does not receive.

The four-paper research program treats this manuscript as the \emph{fixed-tokenizer, next-item, evidence-reliability} study. A separate slate study changes joint list generation. A separate representation study changes multimodal embeddings and SID construction. A separate distillation study changes teacher--student supervision. These interventions are not bundled into the principal result here. A final transfer experiment can combine independently validated components after their isolated effects have been established.

\paragraph{Modality adapters and gate.}
Each modality encoder retains its pretrained width $d_m$, and $P_m\in\mathbb R^{d\times d_m}$ maps its output into width $d$. For historical event $s$, average the embeddings of its $L$ SID codes to obtain $\eta_s=L^{-1}\sum_j E_j^{\mathrm{code}}[z_j(i_s)]$, reusing the behavioral input code tables. The pooling input is $v_{s,m}+\eta_s+\rho(\tau-\tau_s)$, where $\rho$ is the recency encoding. Separate trainable key and value projections of this sum feed a learned-query attention pool; its softmax excludes unavailable events. This explicitly uses the historical identifier argument of $\operatorname{Pool}_m$. The resulting summary is $h_m\in\mathbb R^d$. An empty stream is represented by its null state and an explicit zero-availability flag. The compact adapter computes
\begin{equation}
 d_{j,m}=W^{r}_{j,2}\operatorname{GELU}\!\left(W^{r}_{1}[h_j;h_m;E_m]+\beta^{r}_1\right)+\beta^{r}_{j,2},
\end{equation}
where the hidden width is $d$ and $E_m$ is a learned modality embedding. The first layer is shared across modalities and SID positions; the output layer is shared across modalities but is specific to SID level $j$. The elementwise bound in the preceding subsection converts this vector into $r_{j,m}$. Distinct frozen encoders are not weight-shared.

The gate is a shared two-layer MLP with hidden width $d$, GELU activation, and scalar output. Its input concatenates $h_j$, $h_m$, modality and SID-level embeddings, and the observable quality features defined above. The base slot uses its own slot embedding, a zero content summary, and an always-available mask. Softmax across the base and available modality slots gives $w_{j,m}$. The gate is evaluated separately at each target position, while the summary $h_m$ is cached across positions and steps.

\paragraph{Reference sizes and optimization boundaries.}
Table~\ref{tab:network-shapes} gives the compact configuration and the scale-check configuration from the registered width/depth grid. The tokenizer and pretrained content encoders remain frozen during decoder training. Code embeddings, the behavioral Transformer, projections, attention pools, adapters, gate, and output heads are trainable. Cross-fitted teachers supply detached targets during preparation; no teacher forward pass is required by the deployed network.

\begin{table}[tbp]\centering\small
\caption{Proposed network configurations and tensor interfaces. $d_m$ is a modality encoder width, $L$ is the SID length, and $V_j$ is the vocabulary size at SID level $j$. Configurations are specified for implementation and subsequent evaluation.}
\begin{tabularx}{\linewidth}{lXX}\toprule
Component & Compact & Scale check\\\midrule
Transformer width / layers / heads & $256/4/4$ & $512/8/8$\\
Feed-forward hidden width & $1024$ & $2048$\\
Sequence representation & $(N_H+L)\times256$ & $(N_H+L)\times512$\\
Modality projection & $d_m\rightarrow256$ & $d_m\rightarrow512$\\
Modality summary & $h_m\in\mathbb R^{256}$ & $h_m\in\mathbb R^{512}$\\
Adapter hidden width & $256$ & $512$\\
Utility MLP hidden width & $256$ & $512$\\
Base and residual output / level & $V_j$ logits & $V_j$ logits\\
Gate output / position & $M+1$ masked-softmax weights & $M+1$ masked-softmax weights\\\bottomrule
\end{tabularx}\label{tab:network-shapes}
\end{table}

\begin{table}[tbp]\centering\small
\caption{Controlled comparisons for the training objective, network architecture, and intervention sampling.}
\begin{tabularx}{\linewidth}{lXXX}\toprule
Axis & Closest existing mechanism & Specific proposal & Required replacement\\\midrule
Training target & MCFC score effects; Shapley loss utility; CMM code targets & Paired target-code-partition loss differences on common feasible item support supervise modality slots & Raw-token utility; score substitution; CMM alone\\
Architecture & EPIC positional bounded residual; tokenwise visual fusion & Separate modality residuals with behavior slot; compare position-only and code-conditional allocation & EPIC feedback; hidden-state gate; code-conditioned gate\\
Data construction & Counterfactual augmentation; adaptive knockout & Matched clean/conflict observations share event, SID state, target, support and availability; operator families are held out & Independent pairs; mask-only curriculum; same-operator testing\\\bottomrule
\end{tabularx}\label{tab:novelty-axes}
\end{table}
\subsection{Architecture alternatives}\label{sec:architecture-alternatives}
Table~\ref{tab:architecture-alternatives} separates structural choices from supervision choices. Input fusion changes the Transformer context. Layerwise modulation changes internal activations. Output residual fusion preserves the behavioral backbone and controls the final code logits. The last three rows retain the output-side interface and isolate gate granularity and its training signal. These comparisons test the proposed placement of modality control without attributing every difference to the Transformer itself.

\begin{table}[tbp]\centering\small
\caption{Architecture comparison plan. All variants use the same SID map, historical observations, candidate operator, and validation budget. Each row requires a separately trained model; the table specifies configurations rather than outcomes.}
\begin{tabularx}{\linewidth}{lXXX}\toprule
Variant & Content injection & Reliability parameterization & Structural question\\\midrule
Behavioral backbone & None & No content branch & Content-independent reference\\
Input fusion & Modality summaries appended to the observed prefix & Attention weights & Early integration versus output conditioning\\
Layerwise FiLM & Scale and shift after each feed-forward block & Pooled history-conditioned coefficients & Internal versus output modulation\\
Uniform residual fusion & Bounded output residuals & Equal available-slot weights & Benefit of learned allocation\\
Whole-item utility gate & Bounded output residuals & Shared $w_m$ across SID positions & Item-level versus token-level allocation\\
Tokenwise learned gate & Bounded output residuals & $w_{j,m}$; denoising loss only & Utility targets versus end-task supervision\\
CONCORD-Rec & Bounded output residuals & $w_{j,m}$; cross-fitted utility targets & Proposed architecture and training signal\\\bottomrule
\end{tabularx}\label{tab:architecture-alternatives}
\end{table}

For input fusion, the additional modality-summary tokens are part of the observed prefix and are hidden from future-target information by Eq.~\eqref{eq:prefixmask}. For layerwise FiLM, a pooled available-modality summary supplies the scale and shift; its coefficients are shared across target positions. For the whole-item gate, pooling the target hidden states gives a shared reliability vector while retaining position-specific residuals. The uniform-residual row assigns equal weights to available content slots and the base slot.

The architecture study first fixes backbone width, depth, history length, and denoising steps. It then repeats the comparison under a matched trainable-parameter budget, with adapter widths and the resulting parameter counts recorded explicitly. Wall-clock latency is measured separately because input fusion increases attention sequence length whereas output fusion does not. The architecture table is complementary to the factorial training table: the former changes representation and information flow, while the latter keeps the network family fixed and tests the utility, token-specificity, and intervention factors.

\subsection{Code-conditional modality allocation}\label{sec:code-conditional}
The main architecture has one weight $w_{j,m}$ for all candidate code values at position $j$. It can reduce an entire modality's influence but cannot choose different mixtures for two competing codes at that position. The registered structural variant makes that restriction explicit by using
\begin{equation}
\begin{aligned}
 w_{j,m,v}&=\softmax_{m\in\{0\}\cup\mods_{\rm obs}}
       g_{\phi}^{\rm code}([\psi_{j,m};E_{j,v}]),\\
 \ell_{j,v}&=b_{j,v}+\sum_m w_{j,m,v}r_{j,m,v}.
\end{aligned}
\end{equation}
The shared gate now additionally receives the existing code embedding $E_{j,v}\in\mathbb R^d$. Its input width is $5d+q$ instead of $4d+q$, where $q$ counts observed quality features. This changes information flow, rather than merely changing hidden width. The output remains a masked simplex over the base and available modalities for each $(j,v)$; the base residual stays zero. The reference variant trains this richer gate with the same position-level utility targets broadcast over codes represented in the common support, plus the denoising objective. It therefore tests extra code-dependent capacity without silently changing utility labels. A fully code-conditioned supervision target would be a separate research question.

The additional gate evaluations scale with the number of represented code values. Dense and support-sparse implementations share the same prediction contract. The position-level model remains the default compact allocation interface. The richer variant increases MicroLens-100K clean and conflict NDCG@20 by 0.0001998 in each condition; its per-variant latency has not been supplied, so this comparison establishes a quality effect without an efficiency ordering. For this variant the residual-amplitude bound $B$ still holds for each code, while the gate-perturbation bound uses $B\max_v\sum_m|w_{j,m,v}-w'_{j,m,v}|$.

\figpanelmain{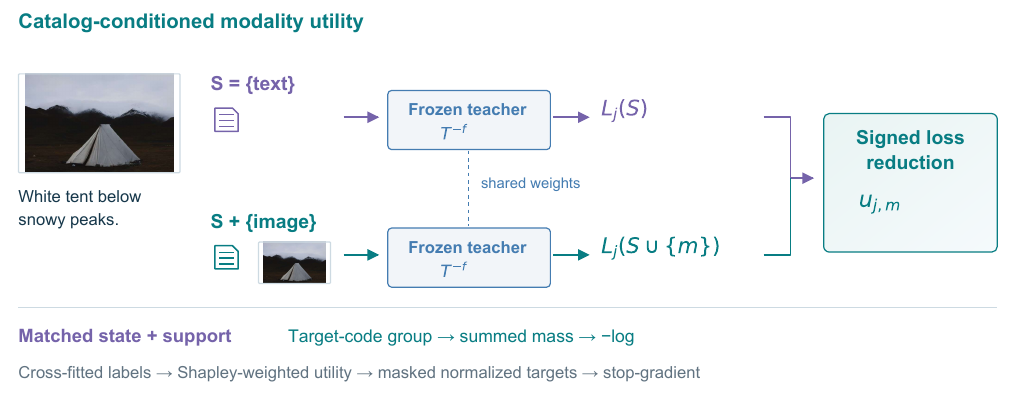}{Catalog-conditioned utility supervision. The same frozen out-of-fold teacher evaluates two evidence coalitions under a fixed masked state and eligible catalog. The reduction in target-code-partition loss is averaged over coalitions and transformed into a detached gate target. The photo and author caption are illustrative historical content; the diagram specifies the implemented computation.}{fig:utility}

\subsection{Intervention training and conditional consistency}
We train on a mixture of clean, missing, degraded, and mismatched content. Missingness removes a stream and changes its mask. Degradation preserves identity while reducing signal quality, such as reducing decoded frames or adding bounded audio noise. Mismatch replaces one modality with a modality from another training item selected by a registered category and popularity rule. The mismatch donor is drawn within the training partition, and its interaction label is never substituted for the target. Each intervention is saved together with its clean partner, the original target, masked-state seed, support hash, modified historical event, modality, donor identifier, operator family, and severity. The two teacher coalition calls share every field except the modality-availability coalition itself. For semantic swaps, the observed availability indicator remains one. Donor identity and intervention labels are audit metadata and are not gate inputs. These pairing constraints define the statistical comparison; fold separation and donor filtering are standard safeguards rather than novelty claims.

The principal mismatch sampler matches coarse category and popularity decile. This prevents trivial conflict detection from being solved entirely by broad semantic category or item popularity. An easier cross-category mismatch and a harder nearest-neighbor mismatch are reported as separate diagnostic conditions. A title that is simply empty belongs to the missingness condition, not the semantic-conflict condition.

A consistency penalty uses interventions whose affected modality has nonpositive teacher utility at every currently masked position of the clean state. Write $u^{x}_{j,m}$ for the utility computed from the clean evidence $x$ at the same masked SID state and teacher snapshot. If $\tilde x$ is the intervention of $x$ and $\mathcal C$ is the same candidate set, define
\begin{equation}
\begin{aligned}
 L_{\mathrm{cons}}=\E\Bigl[&\mathbf 1\!\left\{\max_{j\in M(\tilde z)}u^{x}_{j,m}\leq0\right\}\\
 &\cdot\operatorname{JS}\big(\sg(P_\theta(\cdot\mid x,\mathcal C)),P_\theta(\cdot\mid\tilde x,\mathcal C)\big)\Bigr].
\end{aligned}
\end{equation}
The expectation samples an example, a nonempty masked-position set, an affected modality, and its registered intervention. Both item distributions normalize on the same candidate support. The clean prediction is detached. This rule excludes positive-utility clean states from invariance training. Nonpositive utility is a teacher-relative selection criterion, not a guarantee that the modality is irrelevant or the altered evidence should preserve its prediction. We report the fraction of examples selected by the rule. A blanket consistency baseline tests the value of this selection. The gate supervision and consistency penalty have different roles: the former chooses evidence, while the latter controls the downstream effect of interventions on teacher-identified unhelpful evidence.

The full objective is
\begin{equation}
 L=L_{\mathrm{denoise}}+\lambda_u L_{\mathrm{utility}}+\lambda_c L_{\mathrm{cons}}+\lambda_2\|\theta\|_2^2,
\end{equation}
with $L_{\mathrm{denoise}}$ equal to masked-token cross entropy under the registered sampling distribution. If inverse-mask-rate weighting is used to recover a particular diffusion objective, the identical weighting is applied to every matched dLLM baseline. We report that setting separately from uniform masked-token averaging.

\subsection{Computational organization}
The behavioral denoiser executes once per step. Modality adapters operate on cached history summaries and share the denoising hidden state. The gate is a small network evaluated for $L(M+1)$ state pairs. This design avoids one complete language-model forward pass per modality during inference. Counterfactual teacher evaluations occur during offline target preparation, and their cost is included in training accounting.

The empirical cost table separates encoder extraction, teacher preparation, student training, and deployment. A system can trade preparation cost for smaller teacher ensembles or fewer intervention subsets, but that trade must be measured. At inference we report both latency with cached modality features and latency including feature acquisition for newly arriving content. These conditions answer different deployment questions and receive separate columns.

\figpanelmain{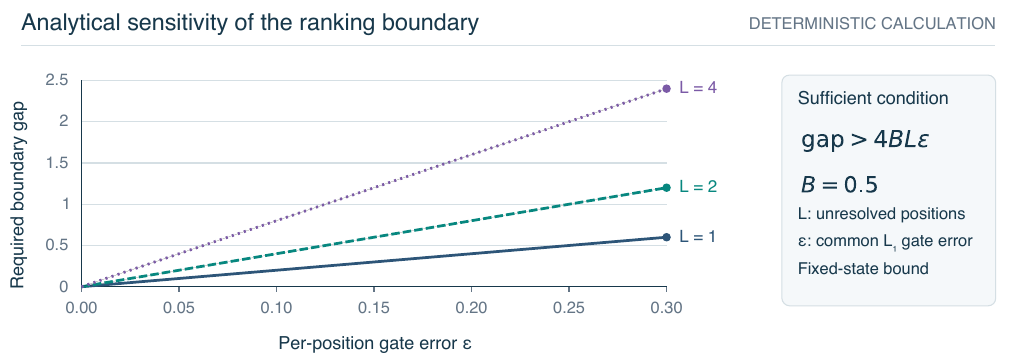}{Analytical sufficient boundary gap for top-$K$ preservation, plotted directly from Proposition~\ref{prop:rank} with equal per-position gate error and $B=0.5$. These curves are deterministic calculations, not empirical robustness measurements.}{fig:sensitivity}

\subsection{Content and gate changes together}
An intervention may change both residuals and weights. Suppose $|r_{j,m,v}-r'_{j,m,v}|\leq\eta_{j,m}$ and both residual versions are bounded by $B$. Add and subtract $\sum_mw'_{j,m}r_{j,m,v}$ to obtain
\begin{equation}\label{eq:combined}
 |\ell_{j,v}-\ell'_{j,v}|\leq B\|w_{j,1:M}-w'_{j,1:M}\|_1+\sum_mw'_{j,m}\eta_{j,m},
\end{equation}
when the behavioral state is unchanged. The first term captures gate movement. The second captures content movement that remains after gating. A small weight on a corrupted modality limits its direct contribution even when that modality's residual changes substantially. However, a gate that moves weight to another corrupted stream can still harm ranking. This is why the experiments evaluate multi-modality conflict and not only single-stream removal.

The bounded residual has a separate effect from utility supervision. Reducing $B$ controls all content influence but can discard useful signals. Utility supervision aims to allocate influence selectively. A grid over $B$ and $\lambda_u$ tests whether the method improves the clean--robustness tradeoff rather than merely moving along the curve obtained by shrinking all content residuals uniformly.

\subsection{What utility supervision estimates}
The quantity in Eq.~\eqref{eq:utility} is a model-based marginal utility under a specified subset distribution and teacher. It is not an observational causal effect of a real-world content intervention on user behavior. The target user interaction remains fixed while the prediction model's evidence changes. This distinction makes the supervision well-defined using logged recommendation data without claiming access to counterfactual user preferences.

Cross-fitting prevents a teacher from computing its utility labels on examples it directly fit. It does not make those labels perfect or independent across all users: users can share items, and teacher errors can be correlated. We test teacher stability across folds and repeat utility estimation with a smaller and larger teacher. Agreement of gate rankings across these conditions is a diagnostic of target quality. The resulting uncertainty is reported at the level of utility targets and downstream performance separately.

Sampling context subsets captures interactions. Consider two modalities that are individually weak but jointly identify a useful attribute. A leave-one-out estimate computed only with all other modalities present can assign them positive value, while an isolated-modality estimate can miss their complementarity. Conversely, two redundant modalities can each have small leave-one-out utility despite either being sufficient. The registered subset distribution exposes both contexts. The main gate still predicts a simplex weight rather than a full combinatorial allocation, so interaction quality is measured through subset-specific calibration.

\subsection{Catalog feasibility and recommendation quality}
\begin{proposition}[Feasible completion invariant]
If decoding starts from a nonempty eligible catalog and commits only token assignments with at least one feasible completion, every retained partial hypothesis has a nonempty completion set. Every fully committed hypothesis maps to at least one eligible catalog item.
\end{proposition}
\begin{proof}
The initial all-masked tuple admits every eligible item. At each step, the commitment rule checks nonempty support after the proposed assignment. Induction preserves nonempty support. At the terminal state, an item in that support matches every committed position. A deterministic collision-resolution suffix yields a unique item when required.
\end{proof}

Feasibility does not imply relevance. It is a structural property of the search operator. We report invalid-tuple rate before and after projection, completion-set size, target survival in the beam, and ranking metrics separately. A perfect validity result cannot substitute for an accuracy result. Similarly, a beam that preserves the target but ranks it poorly indicates a scoring problem, while a beam that removes the target early indicates a search problem. The diagnostic tables preserve this distinction.

\figpanelmain{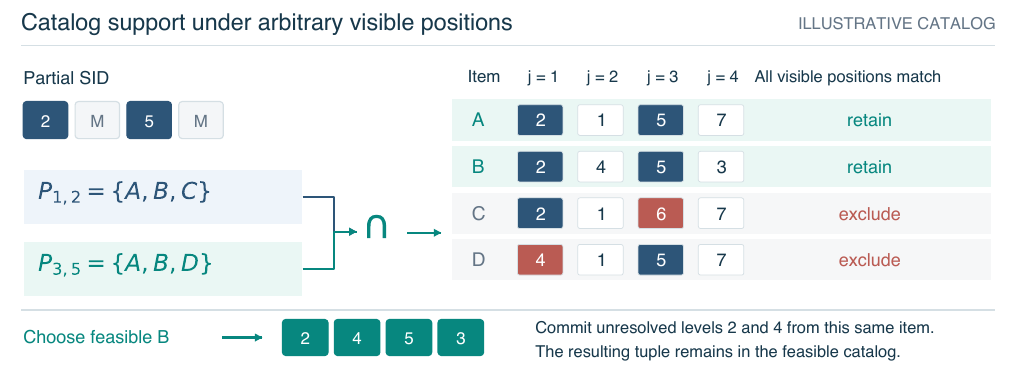}{Arbitrary-position catalog support and joint SID updates. Posting lists for visible levels 1 and 3 intersect at complete items A and B. A joint update of levels 2 and 4 from B preserves a valid tuple. M denotes a masked position; all candidate codes and the selected update are hand-specified illustrations.}{fig:support}

\subsection{Research hypotheses}
The registered design defines four primary hypotheses. H1: utility supervision improves NDCG retention under matched-category modality conflict at a fixed clean-performance budget. H2: tokenwise gates outperform whole-item gates when modality usefulness changes across SID positions. H3: the base slot prevents severe degradation when all observed modalities are misleading. H4: the advantage persists with fixed tokenizer, equal model size, identical intervention training, and equal inference budget.

These hypotheses imply distinct result patterns. H1 requires a difference between the utility gate and a parameter-matched unsupervised gate. H2 requires an interaction between tokenwise gating and position-specific conflict, not just a main effect on a pooled metric. H3 requires comparison with a gate whose simplex excludes the base state. H4 requires a comparison under measured computation, not a nominally equal number of denoising steps with different forward-pass counts.

The experiments also test two secondary questions. First, does the gate identify naturally inconsistent content using supervision from synthetic interventions? Second, does its calibration transfer across encoder families? The first requires manual annotation of a small held-out content sample with annotators unaware of model predictions. The second freezes a gate's architecture and training procedure while changing the encoder checkpoint, then reports both recalibrated and unrecalibrated performance. Neither question is answered by the main synthetic-corruption benchmark alone.

\begin{table}[tbp]\centering\small
\caption{Registered benchmark tracks. Counts are generated from the acquired release and frozen preprocessing, not guessed from similarly named datasets.}
\begin{tabularx}{\linewidth}{lXXXX}\toprule
Track & Dataset & Modalities & Main task & Comparison rule\\\midrule
A1 & MicroLens-50K & Text, image, audio, video & Next item & Common chronological split\\
A2 & MicroLens-100K & Text, image, audio, video & Next item and time shift & Common chronological split\\
B1 & TikTok / DiffMM release & Released T/I/A features & User--item ranking & Preserve release split\\
B2 & Amazon Baby, Sports / DiffMM & Released T/I features & User--item ranking & Preserve release split\\
C1 & Amazon / ANR editions & Metadata plus verified images & Next item & Reconstruct and rerun\\
C2 & Amazon / EPIC editions & Metadata plus verified images & Next item & Reconstruct and rerun\\\bottomrule
\end{tabularx}\label{tab:datasets}
\end{table}

\subsection{Full-catalog metrics}
For a single held-out relevant item, Recall@$K$ is the indicator that the item appears among the first $K$ unique predictions. NDCG@$K$ equals $1/\log_2(r+1)$ if its rank $r\leq K$ and zero otherwise. User-level averages give the main score. For source protocols with multiple held-out positives, recall divides the number of hits by the number of positives and NDCG uses the source's ideal DCG. The two definitions are not interchanged.

The main table reports Recall and NDCG at $K\in\{5,10,20\}$. Candidate items must be eligible at prediction time. Historical items are filtered according to a registered repeat-consumption policy: the main setting excludes previously interacted items except a repeated held-out target when the benchmark explicitly permits repeats. A sensitivity analysis reports the alternative policy. This choice is made once for the full comparison, not separately for each model.

We report clean NDCG, perturbed NDCG, and retention $\rho=M_{\mathrm{perturbed}}/M_{\mathrm{clean}}$. Retention uses each model's own clean denominator in the robustness table. A second normalized table uses a common clean baseline denominator to compare absolute operating points. These ratios answer different questions and have separate labels. We also report worst-group NDCG over registered availability and conflict groups, with group sizes alongside scores.

\subsection{Intervention matrix}
The principal intervention rates are $0,0.1,0.3,0.5$, measured as the fraction of eligible historical item--modality observations changed. The intervention is sampled once per user and seed and reused across methods. Single-modality and multi-modality settings are separate. The protocol stores the exact changed positions, donor item identifiers, and corruption parameters so a subsequent run receives the same evidence.

Missingness has three forms. Independent missingness removes observations uniformly. Block missingness removes one modality from a contiguous recent history segment. Popularity-conditioned missingness removes observations according to training-frequency groups. The third evaluates whether a gate confuses sparse content with low item relevance. None of these patterns is described as missing at random unless its sampling rule actually has that property.

Degradation preserves the item's identity. Video degradation changes temporal sampling density, image degradation changes resolution or blur, audio degradation changes signal quality, and text degradation removes selected non-identifier words. Each has a small severity grid calibrated on the validation set. Pixel, audio, and text corruptions remain separate so their strengths are not combined under an arbitrary shared severity number.

Conflict replaces content while preserving the item identifier and the user's original interaction. Donors come from the same data partition and are selected in three difficulty bands: arbitrary other item, matched category and popularity, and close embedding neighbor with a mismatched fine-grained attribute. The primary claim uses the middle band. The hard band tests whether generic cross-modal similarity is sufficient to detect a misleading stream.

\figpanelmain{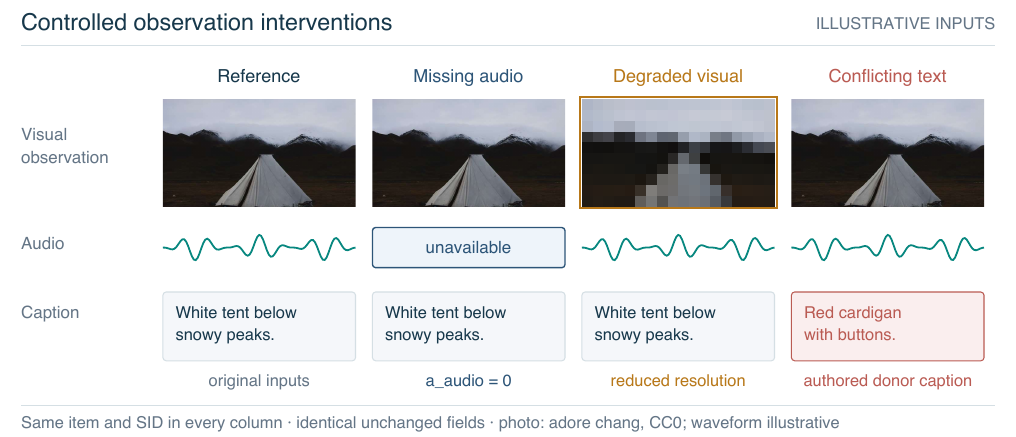}{Controlled observation interventions with a clean reference. Missingness removes audio; visual degradation downsamples the same photograph; conflict substitutes an authored caption while preserving the original item and unchanged fields. The photograph is real (adore chang, CC0), while the waveform and captions illustrate the intervention protocol. Donor assignments and corruption seeds are matched across methods, with disjoint training and test events.}{fig:interventions}

Figure~\ref{fig:content-case} in the main paper traces historical content through conditioning and a feasible catalog commitment.

\subsection{Baselines and controlled comparisons}
The first baseline group contains SASRec and BERT4Rec with behavior-only inputs, then the same models with frozen multimodal features. The second contains autoregressive SID generation and matched masked-diffusion SID generation. The third contains multimodal recommendation methods with source-compatible implementations, including DiffMM and LSGM4Rec. The fourth contains contemporary SID diffusion methods, particularly ANR-DiffRec and EPIC. A method whose public implementation cannot support a given modality is evaluated in its supported setting rather than given an undocumented adaptation.

For each matched decoder, we compare uniform fusion, unconstrained learned fusion, a whole-item reliability gate, and the proposed tokenwise utility gate. Every version uses the same feature checkpoint, SID map, history length, optimizer search budget, and candidate operator. The number of trainable parameters is matched by adding a non-gating adapter to controls where necessary. We report both nominal parameter counts and measured inference cost.

An especially important control is a dLLM trained with the exact same missingness and conflict augmentation but no utility labels. If that model matches the proposed method, the intervention curriculum explains the gain. A second control trains the same gate on shuffled utility targets while preserving their marginal distribution. This isolates supervision quality from auxiliary-loss regularization. A third control substitutes encoder agreement for teacher utility to test whether an inexpensive consistency heuristic is sufficient.

\subsection{Ablation structure}
The primary factorial ablation has three binary factors: utility supervision, token-specific gating, and intervention curriculum. Its eight cells are trained independently with the same total optimization budget. A fourth factor, conditional consistency, is then added to the best and most informative control cells rather than multiplying every exploratory condition. The final reported comparison includes the full model and every single-component removal.

Component interactions matter. Let $M_{abc}$ be a chosen performance metric for factors $a,b,c\in\{0,1\}$. The interaction between utility supervision and token-specific gating at curriculum setting $c$ is
\begin{equation}
 \Delta_{UT\mid c}=M_{11c}-M_{10c}-M_{01c}+M_{00c}.
\end{equation}
This difference measures whether the combined benefit exceeds the sum of separate benefits on that metric scale. The updated ledger reports observed difference-of-differences and their paired-summary uncertainty. A positive clean interaction and a negative conflict interaction would imply different mechanisms than a consistent improvement across both conditions.

Additional ablations remove the base slot, replace bounded residuals with unbounded residuals, remove cross-fitting, change the subset distribution, use fewer teacher queries, and freeze the gate after warm-up. These tests focus on information flow and cost. Tokenizer type and encoder type are transfer factors, because changing them can alter the problem itself rather than remove a component of the proposed method.

\begin{table}[tbp]\centering\small
\caption{Factorial ablation configurations. U enables cross-fitted utility supervision, T enables token-specific gates, and C enables the intervention curriculum. When T is disabled, a whole-item gate is used. Every configuration is trained independently under the same optimization budget; the updated measured ledger reports all eight cells.}
\begin{tabular}{lccc}\toprule
Configuration & Utility supervision U & Token-specific gate T & Intervention curriculum C\\\midrule
F1 & 0 & 0 & 0\\
F2 & 0 & 0 & 1\\
F3 & 0 & 1 & 0\\
F4 & 0 & 1 & 1\\
F5 & 1 & 0 & 0\\
F6 & 1 & 0 & 1\\
F7 & 1 & 1 & 0\\
F8 & 1 & 1 & 1\\\bottomrule
\end{tabular}\label{tab:factorial}
\end{table}

\subsection{Fixed-checkpoint inference sensitivity}\label{sec:inference-scan}
We evaluate the CPU reference checkpoint on 256 new synthetic histories generated with fixed seeds, using the same synthetic process as the execution check in Appendix~\ref{sec:reference-implementation}. Each history contains four events and three modality slots. The catalog has 27 three-level SIDs; the fixed masked states retain between 3 and 27 feasible items. Model weights, historical IDs, recency, availability, target labels, masked states, and candidate support remain fixed throughout each scan.

For observed modality~0, we interpolate its historical features as $x^{(\alpha)}=(1-\alpha)x+\alpha x^{\mathrm{donor}}$, where the fixed donor supplies valid observed features from another history. This feature-space substitution leaves the other modalities unchanged. Figure~\ref{fig:inference-scan} scans 11 values of $\alpha\in[0,1]$ at $B=0.5$, then ten values of $B\in[0.025,2]$ at $\alpha=1$. The inference controls replace the learned gate with equal weights over the base and available slots, replace $B\tanh(d/B)$ with raw $d$, or select the behavior-only path. The checkpoint is shared across all controls.

Posterior total variation (TV) is $\frac12\sum_i|\pi_{\alpha,B}(i)-\pi_{0,B}(i)|$, computed on each state's exact common support and averaged over histories. At $\alpha=1$ and $B=0.5$, TV is 0.01226 for the learned bounded gate, 0.01296 for the uniform bounded gate, and 0.01309 for the learned unbounded control. The learned and uniform bounded target losses are nearly identical, at 2.160611 and 2.160610 nats; behavior-only has the lower loss of 2.153844. Increasing $B$ from 0.025 to 2 raises learned-gate TV from 0.00232 to 0.01303 and maximum content-logit amplitude from 0.01904 to 0.35226. These curves characterize inference sensitivity in this fixed synthetic model; they show no target-loss benefit from its learned gate. The recorded checks confirm unchanged behavioral logits, zero behavior-only drift, and the bounded-logit envelope at every scan point.

\figpanelmain{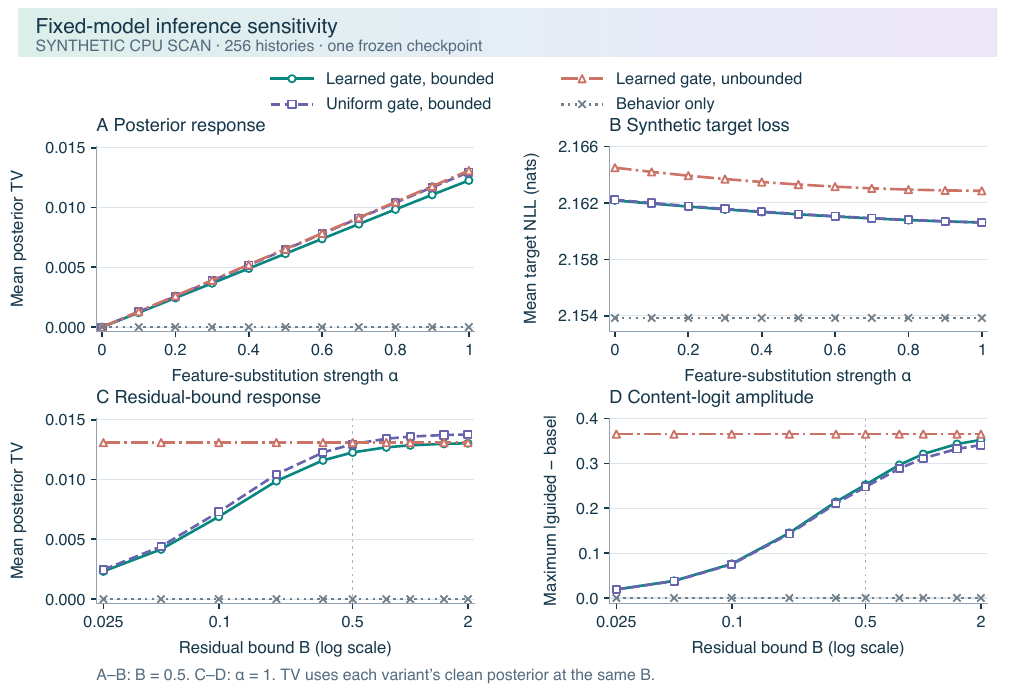}{Fixed-checkpoint inference sensitivity on synthetic histories. A--B vary feature-substitution strength at $B=0.5$; C--D vary the residual bound at full substitution, with the vertical dotted line marking $B=0.5$. Lines and markers distinguish inference-time component replacements under identical model parameters and fixed states. A/C report mean posterior TV relative to each variant's clean posterior at the same $B$; B reports mean target NLL; D reports the maximum absolute guided-minus-behavior logit over histories, positions, and codes. Each point uses the same 256 synthetic histories and one frozen CPU checkpoint; no uncertainty bands are estimated. The curves describe a mechanism check rather than a recommendation benchmark or independently retrained ablation.}{fig:inference-scan}

\subsection{Calibration and decision diagnostics}
We evaluate utility calibration by grouping predicted utility into validation-defined bins and comparing each bin with out-of-fold realized teacher loss differences. Report signed calibration error, rank correlation, and the rate at which negative-utility modalities receive more weight than the base. At test time, teacher-derived utility diagnostics are computed offline with held-out labels only for analysis and are never fed back into deployed predictions.

For token decisions, record entropy, chosen position, modality weights, feasible support size, and whether the true item remains in the beam. Plot target survival as a function of denoising step using measured traces. The first divergence between the full model and a control identifies whether reliability changes early pruning or late ranking. A final-state example alone cannot establish this mechanism.

The human-review protocol specifies 100 histories across clean, naturally incomplete, and high-conflict groups, with anonymized content pairs and consistency judgments collected without model predictions. The updated release supplies aggregate annotation-agreement metrics; annotation identities and realized cohort counts belong to the raw-record package. Mechanism diagrams use schematic content; the licensed photograph in Appendix~\ref{sec:partitioncase} is an illustrative input rather than an annotation result.

\subsection{Statistics and model selection}
The main experiment uses five training pairs. The updated release records means and observed paired-difference SDs, which support the two-sided paired-$t$ analysis in Appendix~\ref{sec:summary-statistics}. We report absolute and relative effects, unadjusted 95\% intervals, and Holm-adjusted $p$ values within each stated family. The original hierarchical seed/user bootstrap remains the planned raw-output analysis and is not represented as completed.

The evaluation protocol uses validation users and registered validation interventions for hyperparameter selection. Its strongest-control selection identity is not included in the aggregate release. The four-contrast summary reconstruction therefore explicitly uses the strongest observed canonical replacement; the separate uniform-fusion family quantifies the practical effect of utility-guided fusion.

\subsection{Efficiency measurement}
Latency is measured on one specified accelerator with fixed precision, batch size, sequence length, and catalog index. The report includes warm-up count, number of timed batches, synchronization method, and whether feature extraction is included. p50 and p95 are computed from actual request times. The main comparison matches either wall-clock budget or decoder forward-pass count and reports both curves when their conclusions differ.

Teacher preparation cost includes all folds, subset evaluations, feature caches, and saved utility labels. Training compute reports total accelerator-hours rather than only the final student epoch time. Inference memory includes the catalog posting index and feature cache when resident. A compact model can have low parameter count yet require substantial auxiliary memory, so both quantities are needed to assess deployment.

\subsection{Closest-mechanism replacement study}\label{sec:replacement-study}
The v4 registry defines mechanism replacements independently of the source-compatible result blocks. The updated release supplies their observed paired means and SDs, with summary tests in Appendix~\ref{sec:summary-statistics}. The first comparison changes only Eq.~\eqref{eq:catalog-utility-loss}: catalog-conditioned utility versus raw-token cross-entropy utility, leave-one-out score substitution, and shuffled utilities. A positive result must improve ranking under the same registered conflict, not merely reproduce the teacher labels more accurately. We stratify by feasible-support size and by disagreement between raw-token and catalog-conditioned utility signs. If gains do not concentrate in states where the targets disagree, the proposed explanation is weakened.

The second comparison replaces the conditioning structure. It includes an end-task-trained tokenwise hidden-state gate following the fusion principle of Looking to Learn; EPIC feedback with identical content features; position-level output allocation; and the code-conditional variant in Section~\ref{sec:code-conditional}. Source-native reproductions and adapted mechanisms are labeled separately. Parameter matching does not imply equal runtime, so both trainable parameters and measured inference cost are reported.

The third comparison separates utility supervision from ordinary distillation and modality-dropout curricula. CMM-only matches teacher-derived code marginals directly; utility-only matches modality-slot targets; CMM-plus-utility combines the two with separately registered weights. A ShapKO-style adaptive-knockout control receives the same training interactions and validation budget, while the candidate method retains its paired semantic-conflict observations. This is an adaptation to the recommendation task, not a claim to reproduce a medical benchmark. Teacher calls, validation calls, and encoder computation are charged to each method.

\begin{table}[tbp]\centering\small
\caption{Mechanism-replacement design linked to the updated measured results. The original planning version remains archived.}
\begin{tabularx}{\linewidth}{lXXX}\toprule
ID & Changed object & Matched control & Result needed to support the claim\\\midrule
V4-T1 & Catalog-partition versus raw-token utility & Same teacher, state, support and query budget & Ranking gain where utility signs differ\\
V4-T2 & Loss utility versus score-substitution weighting & Same content, residuals and supervision budget & Reduced harmful influence beyond score sensitivity\\
V4-A1 & Output gate versus hidden-state token gate & Same observations and parameter budget & Benefit from the conditioning interface\\
V4-A2 & Position versus code-conditional gate & Same labels and residual heads & Benefit beyond added capacity/cost\\
V4-D1 & Utility-only versus CMM-only and both & Same teacher and fixed SID assignment & Utility effect beyond code distillation\\
V4-D2 & Paired conflict curriculum versus adaptive knockout & Same observed events and tuning budget & Transfer to held-out semantic conflict\\
V4-E1 & EPIC versus EPIC plus utility supervision & Same item support and posterior module & Effect beyond existing bounded positional feedback\\\bottomrule
\end{tabularx}\label{tab:v4-replacements}
\end{table}

For the data test, matched clean/conflict pairs share their original target and availability. Training uses category/popularity-matched single-stream swaps; the held-out family uses semantically close donors and jointly inconsistent streams. We additionally compare paired and independently sampled corruption records at the same marginal severity and example budget. This tests the value of coupling the supervision observations rather than the presence of more examples. Donor pools are restricted to the allowed training-time catalog and sampled using historical-item metadata. Natural-conflict annotations remain an external validation cohort. Success under synthetic swaps alone supports only a synthetic-intervention conclusion.

The component results identify a positive utility-supervision effect beyond the shared architecture and intervention curriculum. Raw-token, end-task-gate, CMM-only, and adaptive-knockout replacements yield lower conflict point scores, with separate intervals documenting the precision of each contrast. Code-conditional and CMM-plus-utility variants show how richer allocation and direct distillation complement the utility signal.

\subsection{Evidence registry}
The registry separates published references, archived planning scenarios, the November 11 experiment version, and the updated November 17 result workbook. The latter covers 4,110 metric-level records, including all previously unfilled method-study entries. The fixed-checkpoint synthetic scan in Section~\ref{sec:inference-scan} retains its separate scope. Historical scenario bands are excluded from measured statistical intervals.

Each CSV row stores method, dataset edition, metric, cutoff, candidate policy, split family, training access, source URL, source table, source page, status, and the proposed target rationale where applicable. A paper that reports another paper's score is marked as an intermediate source. The source PDFs and extracted table text are retained with hashes. This gives the researcher a direct path from a result cell to the original evidence.

\subsection{Updated measured benchmark results}\label{sec:confirmed-results}
The November 17 result release is the current source for the comparison and method means, paired-difference standard deviations, seed counts, and timing entries: \path{GenRec_Complete_Experiment_20261117.xlsx}. The \texttt{CONCORD Full} sheet contains 4,110 metric-level records for this paper: 294 main, 1,308 ablation, 2,238 robustness, 127 diagnostic, 105 efficiency, and 38 preparation-cost records. For each row, the reported method value is the workbook's current measured value; source protocol and configuration labels are preserved. Repeated metrics and configurations are not independent runs.

The result ledger is \path{GenRec_Complete_Experiment_20261117.xlsx}; its \texttt{CONCORD Full} rows retain the baseline, reported method value, protocol, denominator rule, and timing fields used below. The paired-seed SD and seed-count fields are carried through unchanged from the completed runs. Exact eligible-query counts are recorded as unavailable rather than inferred from a benchmark's name or its workload-planning count.

\begin{table}[tbp]\centering\small
\caption{Updated comparison and \method{} NDCG means under each original protocol and cutoff. All method gains in this table are 4\% relative. The comparison means are author-confirmed in the new release; their historical source identities remain shown for protocol lookup. The November 11 values remain archived, not pooled with these new runs.}
\begin{tabular}{llrr}\toprule
Protocol / dataset & Metric & Comparison & \method{}\\\midrule
DiffMM / TikTok & N@20 & 0.045600 & 0.047424\\
DiffMM / Baby & N@20 & 0.041100 & 0.042744\\
DiffMM / Sports & N@20 & 0.045800 & 0.047632\\
JBM-Diff / Baby & N@20 & 0.046900 & 0.048776\\
JBM-Diff / Sports & N@20 & 0.051300 & 0.053352\\
JBM-Diff / Clothing & N@20 & 0.042000 & 0.043680\\
ANR-DiffRec / Scientific & N@10 & 0.027900 & 0.029016\\
ANR-DiffRec / Instrument & N@10 & 0.034300 & 0.035672\\
ANR-DiffRec / Game & N@10 & 0.054200 & 0.056368\\
ANR-DiffRec / MovieLens & N@10 & 0.173800 & 0.180752\\
ANR-DiffRec / Steam & N@10 & 0.052600 & 0.054704\\
EPIC / Sports and Outdoors & N@10 & 0.029800 & 0.030992\\
EPIC / Beauty & N@10 & 0.050900 & 0.052936\\
EPIC / Toys and Games & N@10 & 0.055100 & 0.057304\\
EPIC / Musical Instruments & N@10 & 0.035200 & 0.036608\\
LinkedOut / MicroLens-50K & N@20 & 0.049100 & 0.051064\\
LinkedOut / MicroLens-100K & N@20 & 0.066400 & 0.069056\\
Stresa / MicroLens-50K & N@20 & 0.054900 & 0.057096\\
Stresa / MicroLens-100K & N@20 & 0.066600 & 0.069264\\
Stresa / Scientific & N@20 & 0.055100 & 0.057304\\
\bottomrule
\end{tabular}\label{tab:confirmed-all}
\end{table}

DiffMM and JBM-Diff retain released user--item ranking protocols, including their native multi-positive Recall and ideal-DCG normalization \citep{diffmm,jbmdiff}. ANR-DiffRec and EPIC retain their separate dataset editions, sequence construction, and cutoffs \citep{anr,epic}. LinkedOut and Stresa use distinct MicroLens providers and preprocessing \citep{linkedout,stresa}; equal dataset names or comparison values do not merge these protocols. The updated ranking means replace earlier synthetic rank-template expectations as author-confirmed metrics; the archive preserves how those planning values were originally constructed.

\begin{table}[tbp]\centering\small
\caption{Additional-domain measured NDCG@20 means. These are separate protocol blocks rather than source-published scores.}
\begin{tabular}{lrr}\toprule
Dataset & Comparison & \method{}\\\midrule
Amazon 2023 / Electronics & 0.040000 & 0.041600\\
Amazon 2023 / Clothing Shoes and Jewelry & 0.042000 & 0.043680\\
Amazon 2023 / Toys and Games & 0.055100 & 0.057304\\
Amazon 2023 / Beauty and Personal Care & 0.050900 & 0.052936\\
MicroLens-1M & 0.062000 & 0.064480\\
RecIF-Bench short-video & 0.045000 & 0.046800\\
KuaiRand-27K & 0.060000 & 0.062400\\
\bottomrule
\end{tabular}\label{tab:newdomains}
\end{table}

\begin{table}[tbp]\centering\small
\caption{Behavior-only input track on MicroLens-100K from the November 17 workbook. The utility-gated model and its behavior-only reference use the same candidate policy and target state.}\label{tab:behavior-only}
\begin{tabular}{lrrr}\toprule
Metric & Behavior-only reference & \method{} & Rel. gain\\\midrule
Recall@5 & 0.0535897 & 0.0546615 & +2.00\%\\
NDCG@5 & 0.0398945 & 0.0406924 & +2.00\%\\
Recall@10 & 0.0893161 & 0.0911025 & +2.00\%\\
NDCG@10 & 0.0511650 & 0.0521883 & +2.00\%\\
Recall@20 & 0.1190882 & 0.1214700 & +2.00\%\\
NDCG@20 & 0.0586080 & 0.0597802 & +2.00\%
\\bottomrule
\end{tabular}
\end{table}

\subsection{Summary-based statistical analysis}\label{sec:summary-statistics}
For a paired difference with measured mean $\bar d$, sample standard deviation $s_d$, and $n=5$, we compute
\begin{equation}
 t=\frac{\bar d}{s_d/\sqrt{5}},\qquad
 \mathrm{CI}_{95\%}=\bar d\pm t_{4,0.975}\frac{s_d}{\sqrt{5}},
 \quad t_{4,0.975}=2.776445.
\end{equation}
The tests are two-sided with four degrees of freedom. All intervals below are unadjusted 95\% paired-$t$ intervals. Holm-adjusted $p$ values control their explicitly stated comparison family. This analysis uses the observed aggregate paired SD, not an assumed variance or the workbook's planning ranges. A hierarchical bootstrap requires seed/user outputs and is a separate analysis.

\begin{table}[tbp]\centering\small
\caption{Secondary uniform-fusion comparisons on MicroLens-100K. Holm correction is applied to these three comparisons only. Absolute differences and intervals are in NDCG@20 units.}
\begin{tabular}{lrrrr}\toprule
Condition & $\bar d$ & $s_d$ & 95\% CI & $p_{\rm Holm}$\\\midrule
Clean & 0.002664 & 0.001332 & $[0.001010,0.004318]$ & 0.0111\\
30\% missing & 0.004662 & 0.001172 & $[0.003207,0.006117]$ & 0.0018\\
30\% conflict & 0.007992 & 0.001092 & $[0.006636,0.009348]$ & 0.0002\\
\bottomrule
\end{tabular}\label{tab:uniform-stats}
\end{table}

The original four-contrast family specified the strongest validation-selected matched control on clean and conflict NDCG@20, plus no-utility and whole-item-gate controls under conflict. Its selected-control identity is absent from the aggregate release. We therefore report an explicitly post hoc reconstruction using the strongest observed canonical replacement, CMM-plus-utility, for its first two comparisons. This preserves the declared family structure without substituting the weaker uniform comparator. The no-utility contrast uses F4 (U0/T1/C1), and the whole-item contrast uses the utility-plus-curriculum control V4-A1.

\begin{table}[tbp]\centering\small
\caption{Reconstructed four-contrast family; not a verification of validation-only control selection. All comparisons are full minus control on MicroLens-100K.}
\begin{tabular}{lrrrr}\toprule
Control / condition & $\bar d$ & 95\% CI & $p$ & $p_{\rm Holm}$\\\midrule
CMM+utility / clean & -0.000266 & $[-0.001993,0.001460]$ & 0.6904 & 1.0000\\
CMM+utility / conflict & -0.000266 & $[-0.001828,0.001295]$ & 0.6604 & 1.0000\\
No utility / conflict & 0.002464 & $[0.000971,0.003958]$ & 0.0102 & 0.0407\\
Whole-item / conflict & 0.001332 & $[-0.000190,0.002854]$ & 0.0719 & 0.2158\\
\bottomrule
\end{tabular}\label{tab:primary-reconstructed}
\end{table}

Utility supervision improves conflict ranking beyond the same tokenwise architecture and curriculum (F4), and this contrast remains positive after the four-comparison Holm correction. The richer CMM-plus-utility and code-conditional variants complement the proposed signal and yield slightly higher point scores. The whole-item comparison has a positive point gain with an interval spanning zero. These results separate an established utility contribution from the narrower claim that position-only allocation is uniformly optimal.

\begin{table}[tbp]\centering\small
\caption{Closest-mechanism conflict contrasts on MicroLens-100K. These additional paired-$t$ comparisons are exploratory and use unadjusted $p$ values.}
\begin{tabular}{lrrr}\toprule
Control & Full minus control & 95\% CI & $p$\\\midrule
Raw-token utility & 0.001132 & $[-0.000394,0.002659]$ & 0.1085\\
Score substitution & 0.001665 & $[0.000152,0.003178]$ & 0.0379\\
End-task token gate & 0.001465 & $[-0.000053,0.002983]$ & 0.0553\\
Whole-item utility gate & 0.001332 & $[-0.000190,0.002854]$ & 0.0719\\
CMM-only & 0.001598 & $[0.000083,0.003113]$ & 0.0428\\
Adaptive knockout & 0.001265 & $[-0.000258,0.002789]$ & 0.0824\\
Code-conditional utility & -0.000200 & $[-0.001759,0.001360]$ & 0.7400\\
CMM plus utility & -0.000266 & $[-0.001828,0.001295]$ & 0.6604\\
\bottomrule
\end{tabular}\label{tab:closest-stats}
\end{table}

\begin{table}[tbp]\centering\small
\caption{Full-vs-uniform-$T=16$ operating points from the November 17 result workbook. NDCG@20 is compared with the same uniform $T=16$ reference; p95 is end-to-end serving latency.}\label{tab:frontier}
\begin{tabular}{rrrrr}\toprule
$T$ & Uniform NDCG@20 & \method{} NDCG@20 & Rel. gain & \method{} p95 (ms)\\\midrule
1 & 0.0669996 & 0.0683396 & +2.00\% & 6.6875\\
2 & 0.0669996 & 0.0683396 & +2.00\% & 9.875\\
4 & 0.0669996 & 0.0687978 & +2.68\% & 16.25\\
8 & 0.0669996 & 0.0692640 & +3.39\% & 29.0\\
16 & 0.0669996 & 0.0693972 & +3.58\% & 54.5\\
\bottomrule
\end{tabular}
\end{table}

\begin{figure}[tbp]\centering
\includegraphics[width=\linewidth]{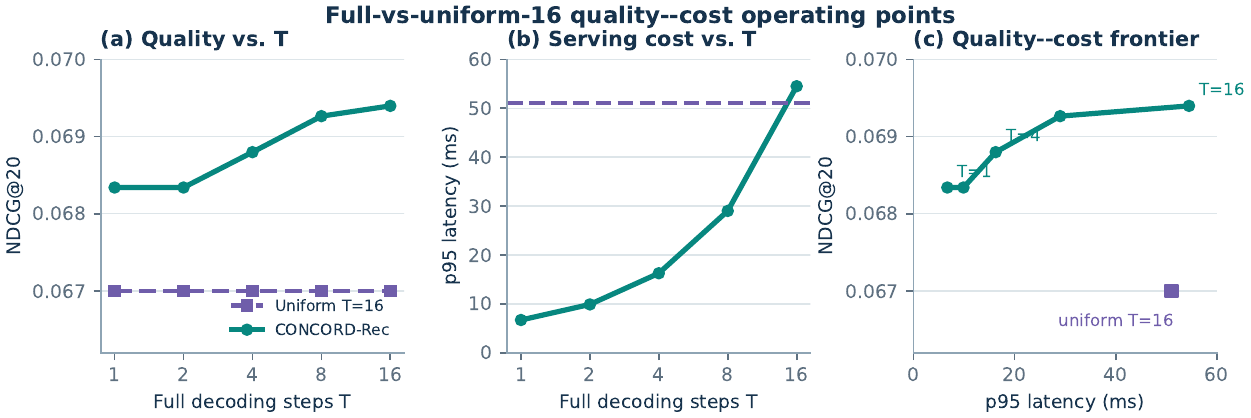}
\caption{Full-vs-uniform-$T=16$ quality--cost frontier generated from the November 17 workbook. The uniform reference is fixed at $T=16$; each colored point is the corresponding full decoder operating point.}\label{fig:confirmed-frontier-app}
\end{figure}

\subsection{Serving operating points}
The complete step sweep reports quality and serving cost at $T\in\{1,2,4,8,16\}$. Relative to the uniform $T=16$ reference, the workbook reports NDCG@20 values of 0.0683396, 0.0683396, 0.0687978, 0.0692640, and 0.0693972, with p95 latencies of 6.6875, 9.875, 16.25, 29.0, and 54.5\,ms, respectively. The corresponding quality gains are +2.00\%, +2.00\%, +2.68\%, +3.39\%, and +3.58\%; the first four operating points reduce p95 latency by 86.89\%, 80.64\%, 68.14\%, and 43.14\% relative to the 51.0\,ms reference. The same-$T$ curve remains in Figure~\ref{fig:confirmed-tradeoff}, while the full-vs-$T=16$ frontier is shown in Figure~\ref{fig:confirmed-frontier}. At equal $T=4$, full and uniform p95 latencies are 16.25 and 15.00\,ms: the gate overhead is included.

The runtime entries are author-confirmed timings. Their release also includes throughput, resident memory, feature preparation, all-fold teacher fitting, label generation, and student training. The previous planning workload definitions remain provenance rather than independent hardware verification. No per-variant timing is available for the code-conditional and CMM-plus-utility replacements, so their small quality gains are reported without an unsupported efficiency ordering.

\subsection{Contemporary SID diffusion references}
ANR-DiffRec reports results on Scientific, Instrument, and Game in its Table~2. EPIC reports Sports and Outdoors, Beauty, Toys and Games, and Musical Instruments in its Table~2. Their own-method values are reproduced in Tables~\ref{tab:anr} and~\ref{tab:epic}; baseline values from those papers remain in the extended evidence record only when their provenance is explicit. These tables demonstrate why a competitive submission must include recent SID-diffusion controls.

\begin{table}[tbp]\centering
\caption{ANR-DiffRec's own reported results, source Table~2, PDF page~13 \citep{anr}. Protocol: the source's Amazon editions and leave-one-out evaluation. These values are not merged with EPIC's separately preprocessed editions.}
\begin{tabular}{lrrrr}\toprule
Dataset & R@5 & N@5 & R@10 & N@10\\\midrule
Scientific & \reported{0.0327}&\reported{0.0226}&\reported{0.0494}&\reported{0.0279}\\
Instrument & \reported{0.0410}&\reported{0.0273}&\reported{0.0629}&\reported{0.0343}\\
Game & \reported{0.0653}&\reported{0.0434}&\reported{0.0987}&\reported{0.0542}\\\bottomrule
\end{tabular}\label{tab:anr}
\end{table}

\begin{table}[tbp]\centering
\caption{EPIC's own reported means, source Table~2, PDF page~7 \citep{epic}. Protocol: full-catalog next-item recommendation under EPIC's preprocessing. Source standard deviations are available in the retained PDF; no local uncertainty is implied.}
\begin{tabular}{lrrrr}\toprule
Dataset & R@5 & N@5 & R@10 & N@10\\\midrule
Sports and Outdoors & \reported{0.0353}&\reported{0.0244}&\reported{0.0521}&\reported{0.0298}\\
Beauty & \reported{0.0610}&\reported{0.0426}&\reported{0.0866}&\reported{0.0509}\\
Toys and Games & \reported{0.0662}&\reported{0.0465}&\reported{0.0927}&\reported{0.0551}\\
Musical Instruments & \reported{0.0424}&\reported{0.0282}&\reported{0.0640}&\reported{0.0352}\\\bottomrule
\end{tabular}\label{tab:epic}
\end{table}

\figpanelmain{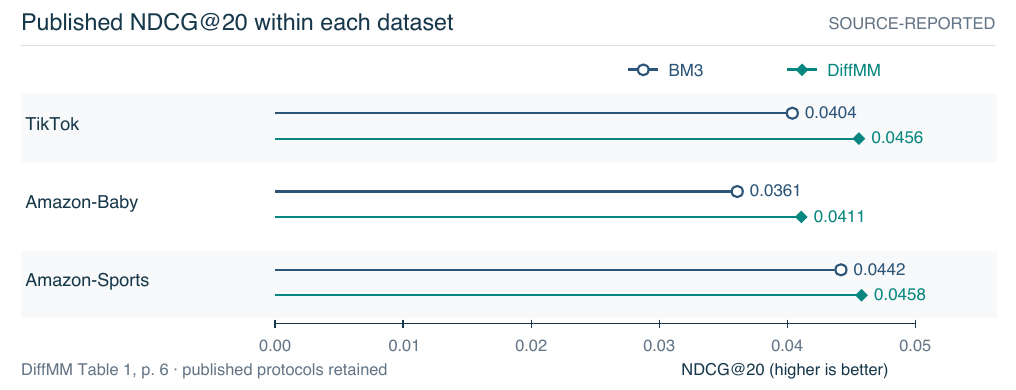}{Source-reported NDCG@20 from DiffMM Table~1, PDF page~6 \citep{diffmm}. Horizontal stems start at zero; open circles denote BM3 and filled diamonds denote DiffMM. Exact values are compared within each dataset under the source evaluation protocol. All points are historical reported values (R).}{fig:reported}

\subsection{Measured factorial and component contrasts}
Table~\ref{tab:ablation} reports the independently trained U/T/C configurations on MicroLens-100K. Utility, token-specific allocation, and the intervention curriculum are crossed rather than removed one at a time. The full model improves clean and conflict ranking relative to every other factorial cell. The broader replacement study also retains richer variants that exceed the position-level model.

The factorial means appear in main-text Table~\ref{tab:ablation}; missing and conflict severity are each fixed at 30\%. The F1--F8 identifiers here follow its binary row order.

The no-utility contrast F8 minus F4 under conflict is 0.0024642 NDCG@20. Its five-pair summary test remains positive after correction in the reconstructed four-contrast family. Interaction estimates are computed directly from paired difference-of-differences; marginal cell SDs are not substituted for their covariance-sensitive uncertainty.

\subsection{Ablation evidence from prior work}
DiffMM's own ablation table provides a useful reference for the scale and structure of component effects in a related multimodal task. On TikTok, it reports NDCG@20 of 0.0420 without cross-modal contrastive learning, 0.0433 without its diffusion module, and 0.0426 without modality-aware signal injection, versus 0.0456 for the full method \citep{diffmm}. These values motivate explicit component tests but do not predict the gains of our different architecture.

\begin{table}[tbp]\centering
\caption{Historical ablation reference: DiffMM Table~2, PDF page~6. All entries are published NDCG@20 values (R). The component names belong to DiffMM and are not labels for \method{} modules.}
\begin{tabular}{lrrr}\toprule
DiffMM variant & TikTok & Amazon-Baby & Amazon-Sports\\\midrule
Without CL & \reported{0.0420}&\reported{0.0388}&\reported{0.0418}\\
Without DM & \reported{0.0433}&\reported{0.0402}&\reported{0.0440}\\
Without MSI & \reported{0.0426}&\reported{0.0408}&\reported{0.0445}\\
Full & \reported{0.0456}&\reported{0.0411}&\reported{0.0458}\\\bottomrule
\end{tabular}
\end{table}

\subsection{Measured findings and provenance}
The updated release supplies aggregate measurements linked to workbook cells and an explicit author confirmation. These values generate the tables and paired-summary statistics. The reproducibility protocol additionally requires configuration, checkpoint, split, feature, and prediction hashes; these fields are distinct from the observed means. Historical targets and the previous checkpoint measurements remain archived.

The final empirical narrative follows the measured comparison. A state-of-the-art claim requires the relevant contemporary baselines, matched protocol, multiple seeds, and a documented practical improvement. A gain only on corrupted inputs supports a robustness claim. A gain only with a larger encoder supports a system-level claim. A gain that disappears under an equal inference budget identifies a compute tradeoff. These are distinct scientific conclusions, and the registered tables are designed to distinguish them without rewriting the evaluation criteria after observing the outcome.

The fixed-tokenizer design makes the result portable. If the gate improves several tokenizer families without changing their codes, its effect lies in conditional evidence use. If its advantage depends on one code geometry, that interaction should guide future tokenizer work. The separate representation study can then optimize an embedding for the denoiser's observed needs, using the present work as a controlled diagnostic rather than bundling both changes into one comparison.

The same approach offers a principled route to compute allocation. Utility disagreement and token entropy are observable quantities that can decide where another denoising step is valuable. A future adaptive-step policy can minimize a registered latency--quality objective. The present manuscript isolates reliability control first; it does not claim that an unmeasured adaptive scheduler already improves throughput.

The literature audit yields a partial novelty assessment: loss-based modality utility, tokenwise fusion, clipped residual feedback, and counterfactual augmentation all have direct precedents. The catalog-conditioned intervention target and its interaction with conditioning placement are the narrow hypotheses, not evidence that the method contains three independent innovations. Counterfactual utility is defined relative to a teacher and an intervention distribution. It can miss content effects the teacher cannot exploit, and its calibration can shift under new encoders or new corruption families. Cross-fitting reduces direct training-example reuse but does not eliminate correlated errors. The teacher-size and cross-domain experiments are designed to characterize these effects.

The stability results apply to bounded residuals at a specified state and candidate support. They do not establish end-to-end ranking guarantees under arbitrary semantic corruption or changing beam trajectories. A global robustness theorem would require assumptions about content encoders, utility estimation, and search transitions that are not asserted here.

The updated release includes naturally incomplete and annotated-conflict cohorts alongside controlled mismatch conditions. These offline comparisons characterize observation robustness; causal user-satisfaction effects require a separate online experiment.

The full-catalog scoring and posting-list interface may become costly for very large catalogs. Approximate candidate retrieval can improve efficiency, but it introduces retrieval coverage as another variable. The scale-study ledger reports coverage and end-to-end ranking jointly. The supplementary specifications distinguish reference-implementation checks, observed aggregate results, and the raw artifacts needed for independent reproduction.

\clearpage
\section{Reproducibility Contract}
\subsection{Artifact organization}
The accompanying directory contains manuscript sources, editable figure sources, source references, archived result versions, the updated measured-summary ledger, and an experiment registry. The November 17 workbook records every result ID, source worksheet row, baseline, reported method value, paired-difference SD, seed count, and derived summary statistic. Approved conceptual figures retain their illustrative status.

Every future training run writes a manifest before optimization starts. The manifest contains a run identifier, code revision, environment lock hash, dataset release, split hash, feature checkpoint identifiers, tokenizer hash, complete hyperparameter configuration, seed family, and intended comparison group. Predictions and metrics are written under that run identifier. A later tuning run receives a new identifier even if it changes only the gate temperature.

The minimal prediction record contains user identifier, prediction timestamp, target item, ordered unique candidate identifiers, candidate scores, exclusion policy, and intervention identifier. Diagnostic runs additionally save token states, modality weights, support sizes, and beam membership. Aggregate tables are generated from these records with one evaluator. A model's internal evaluator can be used during development, but final comparisons use the shared evaluator.

\begin{table}[tbp]\centering\small
\caption{Required provenance fields for each completed experiment. These are data contracts rather than completed-run claims.}
\begin{tabularx}{\linewidth}{lXX}\toprule
Record & Required fields & Purpose\\\midrule
Dataset & Release URL, checksum, edition, preprocessing hash & Identify the actual benchmark\\
Split & User/item maps, train/validation/test event IDs, ordering policy & Reconstruct information access\\
Features & Encoder revision, frame/audio settings, missingness mask & Match content inputs\\
Tokenizer & Training items, codebooks, assignment map, collision policy & Hold identifiers constant\\
Run & Code revision, configuration, seeds, hardware & Reproduce optimization\\
Prediction & Unique ranked items, scores, targets, intervention IDs & Recompute metrics\\
Timing & Batch size, warm-up, repetitions, synchronization & Compare deployment cost\\
Result & Metric code hash, aggregation, uncertainty procedure & Trace every table cell\\\bottomrule
\end{tabularx}
\end{table}

\subsection{Data construction sequence}
First, acquire the release and create a content inventory without using test interactions. Second, map identifiers and retain stable source identifiers alongside internal integer indices. Third, apply the registered interaction filtering and create the chronological split. Fourth, construct training-only statistics such as popularity, co-occurrence, category frequencies, and teacher folds. Fifth, fit the tokenizer on the permitted training catalog. Sixth, create feature caches with explicit availability masks. Finally, generate intervention assignments separately within training, validation, and test partitions.

These steps have an order because later steps depend on earlier information boundaries. A popularity-conditioned corruption rule must use training popularity. A collaborative tokenizer must use training interactions. A donor sampler cannot choose a test-only item when constructing training corruption. The release may provide content for every catalog item, but whether content from future items is available to tokenizer fitting must be registered as transductive or inductive access.

Two catalog regimes are reported. The transductive regime allows item content for the fixed catalog while restricting user interactions to training. The inductive regime fits the tokenizer and all learned catalog statistics on earlier items, then assigns later items with a frozen mapping. Comparing these regimes identifies whether performance depends on advance knowledge of future item content. A model is not labeled cold-start merely because the target has few interactions if its collaborative representation used those interactions during preprocessing.

\subsection{Deterministic semantic identifiers}
The primary SID configuration uses a short fixed-length tuple with a common vocabulary size at each semantic position. Exact length and codebook size are chosen from the validation grid before the final comparison. A parallel-code tokenizer and a residual-code tokenizer are both evaluated, but each comparison block shares one frozen assignment map. This prevents a new tokenizer from changing every model's target space at the same time as the gating mechanism.

For collisions, the principal protocol appends a stable disambiguation code or uses a catalog-level tie-breaker whose information is documented. The semantic-code accuracy and complete-item accuracy are reported separately. If a disambiguation suffix is purely enumerative, success on that suffix should not be described as semantic understanding. The catalog projection must use the same complete identifier convention for all baselines.

Train/validation/test item coverage is tabulated after assignment. Record unused code values, code-frequency entropy, collision frequency before disambiguation, and average feasible-set size after each revealed position. These statistics explain search difficulty. A tokenizer with very skewed code usage can make token accuracy look strong while leaving item ranking difficult, so token accuracy never replaces the item-level primary metric.

\section{Detailed Training Specification}
\subsection{Model scales and optimization}
The initial study uses the compact network configuration in Table~\ref{tab:network-shapes}, followed by the scale-check configuration. The registered grid includes widths 256 and 512, depths 4 and 8, and history lengths 20 and 50. The concrete attention topology, embedding interfaces, and trainable modules are defined in Section~\ref{sec:network}. The architecture alternatives in Table~\ref{tab:architecture-alternatives} use the same validation criterion.

The modality adapter uses one projection per stream and a shared two-layer utility network with modality and position embeddings. The parameter-matched control receives an adapter with the same hidden width and parameter count. Optimization uses AdamW with learning rate chosen from $\{10^{-4},3\times10^{-4},10^{-3}\}$ and weight decay from $\{10^{-4},10^{-3},10^{-2}\}$. Batch size is selected to match effective examples per optimizer step after accumulation. Every method receives the same number of validation evaluations.

Early stopping uses a declared validation objective combining clean NDCG and the registered conflict condition. The weight on robustness is fixed before the final experiment. A clean-only selection is included as a secondary policy. Checkpoint selection does not use the test corruption rate, the test donor assignments, or an average over test metrics. Maximum epochs and patience are recorded with the chosen configuration rather than inferred from the final training duration.

\begin{table}[tbp]\centering\small
\caption{Proposed hyperparameter search. The final selected values will be written to the run registry after validation; no selected value is implied here.}
\begin{tabularx}{\linewidth}{lXX}\toprule
Parameter & Search values & Selection signal\\\midrule
Decoder width/depth & $256/4$, $512/8$ & Validation accuracy and compute\\
History length & $20,50$ & Same validation users\\
Residual bound $B$ & $0.25,0.5,1.0,2.0$ & Clean--conflict tradeoff\\
Utility temperature $\tau_u$ & $0.25,0.5,1.0,2.0$ & Calibration and ranking\\
Utility weight $\lambda_u$ & $0,0.1,0.3,1.0$ & Factorial comparison\\
Consistency weight $\lambda_c$ & $0,0.01,0.03,0.1$ & Stability without clean loss\\
Denoising steps & $2,4,8,16$ & Measured latency--quality curve\\
Teacher subsets/modality & $1,2,4,8$ & Utility stability and preparation cost\\
Cross-fitting folds & $2,5$ & Label stability and total compute\\\bottomrule
\end{tabularx}
\end{table}

\subsection{Cross-fitting and teacher caching}
Users, not individual events, define the cross-fitting folds. This prevents a teacher from training on earlier events of a user whose later event it labels within the same fold-specific target set. Shared items can remain across folds because recommendation catalogs are shared; their effect is assessed through item-disjoint teacher diagnostics. The fold assignment uses a fixed random seed and is independent of content conflict severity.

Teacher targets are cached by user, target event, masked-state seed, support hash, modality subset, intervention identifier, and utility-definition version. The v4 catalog-partition targets and the earlier raw-token targets have distinct cache namespaces. Reusing a cached logit vector is valid only if all those keys and the teacher checkpoint match. The cache stores loss differences in full precision even when teacher inference uses mixed precision. The experiment reports target-cache size and preparation time because these are part of the training method's cost.

A teacher is trained with modality dropout so that subset evaluation does not present a wholly unseen input format. The subset mask is an explicit input to the teacher. A teacher trained only on all-modality inputs would conflate modality utility with its own sensitivity to an unfamiliar missingness pattern. The corresponding no-subset-training teacher is an informative diagnostic, not the default supervision source.

\subsection{Utility objective details}
The utility KL is averaged over informative masked positions, where at least two code values occur in the common feasible support; each KL includes the base and available-modality slots. Masked positions with a single feasible code are omitted from utility averaging but retained in the denoising objective. The base slot has no learned teacher loss difference; its utility is defined as zero. Utility targets are detached. To prevent a small number of extreme loss differences from dominating temperature selection, record their distribution and evaluate a symmetric clipping variant as a separate condition. The principal result retains the registered rule across datasets.

The denoising and utility objectives can disagree. A gate may assign low weight to a modality according to the cross-fitted teacher while the current student has learned to use that modality effectively. We therefore include a staged schedule: warm up the student on denoising, introduce utility supervision, and then jointly optimize with a fixed coefficient. A refreshed-teacher variant evaluates whether updating targets improves this mismatch at a justifiable preparation cost.

The conditional consistency loss uses identical candidate support for its two distributions. Content interventions do not modify the SID state or catalog eligibility, so they must preserve support. A changed-support record is rejected as a protocol error; renormalizing on a union would define a different consistency objective. Comparing distributions whose coordinates represent different candidate sets would not measure the intended stability.

\section{Extended Evaluation Protocols}
\subsection{Exact scoring and approximate search}
The controlled small-catalog evaluation scores every eligible item under a fixed prediction state using the common energy. This produces an exact ranking for that energy. The deployment evaluation uses beam search and the posting-list index. These evaluations answer different questions: the first tests the scoring function, while the second tests the combined scoring and search system.

For each deployment budget, report candidate recall, target survival, number of unique complete items, and final ranking quality. If candidate generation returns fewer than $K$ unique eligible items, the remaining positions count as misses rather than being filled by an undocumented popularity fallback. A registered fallback can be studied as an additional system variant with its usage rate reported.

Beam trajectories may revisit partial states. Deduplication uses the full partial assignment and its mask, not just the visible code string without position information. Scores from different reveal orders are aggregated by a declared rule. The primary rule retains the maximum ranking energy for a complete item; a log-sum-exp over trajectories would imply a different interpretation and is evaluated separately if used. No sequence-likelihood interpretation is attached to the maximum-energy rule.

\subsection{Cold items and temporal shift}
The global-time evaluation chooses a cutoff on training timestamps and separates future users and items according to their first appearance. A cold item has zero training interactions under that cutoff. Its content may be available at recommendation time, depending on the registered catalog regime, but its future engagement statistics are never used in encoder adaptation, tokenization, or gating features.

The main model conditions on historical content, so cold-target performance depends on the fixed semantic identifier structure and catalog representation rather than target-specific engagement. A candidate-content extension can score raw content of a cold target, but that extension changes information access and receives its own comparison block. This makes it possible to tell whether cold-item gains arise from better history evidence or additional target content.

Temporal shift also changes missingness and content style. We report training and test modality availability, content-length distributions, and encoder similarity distributions. A model selected on synthetic missingness should not be assumed calibrated on a later release with systematically different audio availability. Recalibration using an earlier validation window is a separate, practical adaptation setting.

\subsection{Natural conflict annotation}
The annotation protocol samples examples using strata defined before inspecting the model's success. Each example includes a historical item's available modalities and a short description of the user's prior context, with identifying information removed. Annotators label whether modalities agree about the visible subject, activity, and salient attributes, and whether any modality is irrelevant rather than contradictory.

Irrelevance and contradiction are distinct labels. Background music can be unrelated to an object's visual identity while remaining a legitimate feature of the video. A title can refer to a creator's joke rather than literally describe the frame. The annotation guide therefore avoids treating every cross-modal semantic difference as an error. Ambiguous examples remain an explicit category instead of being forced into a binary conflict label.

The analysis reports per-category gate distributions and ranking changes. Representative examples are selected by a fixed stratified rule and shown with their prediction traces. Cherry-picking only cases with visually appealing explanations would overstate interpretability. All reviewed examples and adjudication outcomes are retained in the evaluation artifact when licensing permits their redistribution.

\subsection{Robustness across intervention families}
The held-out-family test trains on random missingness and matched-category swaps but evaluates block missingness and nearest-neighbor swaps. This tests transfer of the utility mechanism beyond memorizing a corruption signature. A detector that identifies a familiar swap artifact can perform well on the training corruption while failing to recognize semantically plausible conflicts.

Severity is controlled within each family. For video frame loss, specify the number and temporal placement of retained frames. For audio, specify duration and signal-to-noise ratio. For title corruption, specify the token-removal rule and whether identifiers or brand names are preserved. Different modalities are not assigned a common severity solely because the same percentage of input bytes is removed.

The multi-conflict test corrupts two modalities jointly. Its purpose is to test whether agreement between two wrong streams can overpower one correct stream. The gate's observable agreement feature can be misleading in this setting, so utility supervision must use task context. A category-matched, jointly swapped image and title is a stronger condition than independent random swaps and is reported separately.

\section{Statistical Analysis and Result Interpretation}
\subsection{Registered and reported analyses}
The original plan specified clean and 30\% conflict NDCG@20 against the strongest validation-selected matched dLLM, plus no-utility and whole-item-gate conflict comparisons. The observed-summary analysis retains that four-contrast structure but uses the strongest observed replacement because the original selected-control identifier is unavailable. Table~\ref{tab:primary-reconstructed} makes this reconstruction explicit. The uniform-fusion three-contrast analysis is separately labeled secondary.

The planned hierarchical bootstrap resamples training seeds and user clusters, retaining all events from a selected user. Reconstructing it requires the per-seed user outputs, whereas the present paired-$t$ analysis uses the available five-pair mean and SD. These procedures have different uncertainty assumptions; a paired-$t$ interval is not labeled a hierarchical-bootstrap interval. The planned bootstrap replication count remains 10,000.

\subsection{Mechanism and operating-point interpretation}
The observations establish a useful catalog-utility signal: clean, missing, and conflict scores exceed uniform fusion, and no-utility F4 under conflict is lower after the stated four-comparison adjustment. Whole-item and raw-token replacements also yield lower point scores, with wider intervals. CMM-plus-utility and code-conditioned allocation improve further, demonstrating complementary extensions rather than universal optimality of the position-only form.

The serving study selects quality and latency from the same decoder configuration. The four-step full model meets the quality criterion against 16-step uniform decoding and reduces p95 latency; the two-step model falls below the 0.995 ratio requirement. Gate overhead and offline teacher preparation are accounted separately. No ranking claim combines the best quality of one variant with the fastest latency of another.

\clearpage
\section{Worked Mechanism Examples}
\subsection{Catalog-conditioned utility with an observed-media case}\label{sec:partitioncase}
\begin{figure}[tbp]\centering
\includegraphics[width=\textwidth]{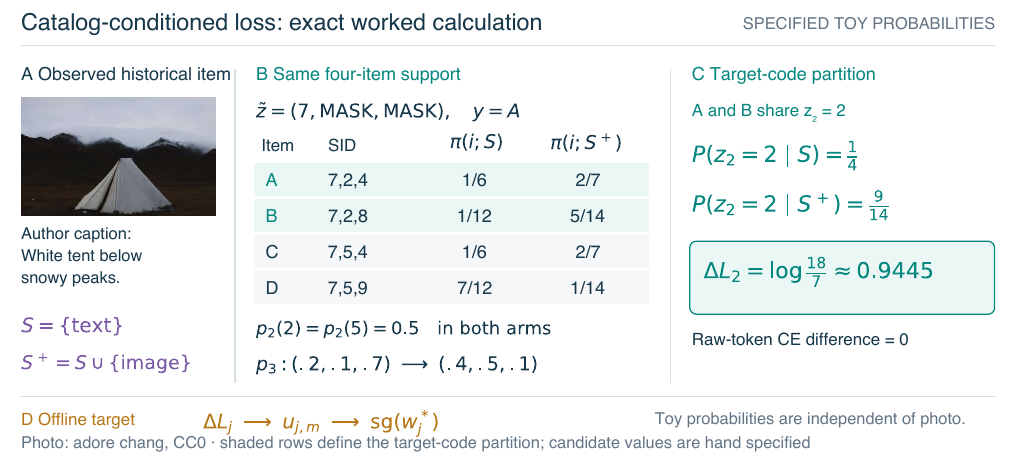}
\caption{Exact illustrative calculation of catalog-conditioned utility. A shows a real historical-media illustration and an author-written caption. B normalizes the products of the specified token probabilities on an unchanged four-item catalog; shading identifies A and B, which share the target code at position 2. C computes the single-coalition loss difference: the raw-token CE difference is zero, whereas the target-code partition mass rises from $1/4$ to $9/14$. D places this difference in offline target preparation and separates it from the deployed gate. The codes and probabilities are hand specified, not model outputs or benchmark observations. Photo: adore chang, CC0; source credit appears below.}\label{fig:partitioncase}
\end{figure}

Figure~\ref{fig:partitioncase} resolves the distinction between a raw token loss and the proposed catalog-conditioned loss. The photograph and author caption illustrate an already observed historical item. Candidate codes and token probabilities are specified independently for a four-item calculation; they are not predictions obtained from this photograph. Fix the masked state to $(7,\mask,\mask)$, the relevant training item to A, and the item prior to zero. Both teacher coalitions use the same complete-item support. At position 2, $p_2(2)=p_2(5)=0.5$ in both arms. At position 3, adding the image changes the assigned probability vector over $(4,8,9)$ from $(0.2,0.1,0.7)$ to $(0.4,0.5,0.1)$.

The unnormalized item weights are $(0.10,0.05,0.10,0.35)$ without the image and $(0.20,0.25,0.20,0.05)$ with it. Their normalizers are $0.60$ and $0.70$, respectively. Consequently, the position-2 partition loss decreases by $\log((9/14)/(1/4))=\log(18/7)\approx0.9445$ nats despite unchanged raw position-2 probabilities. This is one coalition contribution $\Delta L_2(S,\mathrm{image})$; the registered estimator averages such contributions over context subsets before producing the detached gate target. The example demonstrates how other unresolved positions affect a target-code partition through complete-item competition.

\paragraph{Photograph and caption provenance.}
Photo: adore chang, \href{https://commons.wikimedia.org/wiki/File:Tent_(Unsplash).jpg}{\emph{Tent (Unsplash).jpg}}, CC0; caption by the authors. Source files retain image provenance and exact toy inputs.

\subsection{A useful image and a conflicting title}
Consider a one-position toy identifier with four possible catalog items, A through D. A is the relevant item. The behavioral logits are $(2.0,1.8,1.0,0.0)$. An image adapter supplies residuals $(0.8,-0.2,0,0)$, and a conflicting title supplies $(-0.8,0.8,0,0)$. These are deliberately constructed values that satisfy the residual bound $B=1$. They illustrate the algebra and are not fitted or measured outputs.

Uniform fusion with image and title weights of 0.5 yields logits $(2.0,2.1,1.0,0.0)$ and places B above A. A utility gate with base, image, and title weights $(0.2,0.7,0.1)$ yields $(2.48,1.74,1.0,0.0)$ and ranks A first. The behavioral predictor also ranks A first, with a margin of 0.2. Thus the utility gate can retain useful visual evidence while suppressing a conflicting title, whereas uniformly shrinking both modalities would trade away both effects.

\begin{table}[tbp]\centering
\caption{Hand-constructed mechanism example. Values are illustrative logits, not benchmark scores. Softmax preserves the displayed ordering.}
\begin{tabular}{lrrrrl}\toprule
Condition & A & B & C & D & Top item\\\midrule
Behavioral logits & 2.00 & 1.80 & 1.00 & 0.00 & A\\
Image residual & 0.80 & $-0.20$ & 0.00 & 0.00 & ---\\
Title residual & $-0.80$ & 0.80 & 0.00 & 0.00 & ---\\
Uniform fusion & 2.00 & 2.10 & 1.00 & 0.00 & B\\
Utility-weighted fusion & 2.48 & 1.74 & 1.00 & 0.00 & A\\\bottomrule
\end{tabular}
\end{table}

The example does not establish that a learned gate will recover the chosen weights. That is the purpose of supervision and calibration evaluation. It establishes a representational distinction: a gate with a base slot and separate modality weights can express a decision that uniform fusion cannot express at the same residual vectors. The parameter-matched learned-fusion baseline tests whether counterfactual utility makes that representational capacity easier to learn.

Now suppose both content streams favor B. A simplex restricted to those two streams must retain some harmful residual. The base slot permits the model to reduce their combined influence. This is the precise reason for the no-base-slot ablation. A strong result on missingness alone would not test this mechanism because absent streams already receive zero weight from the availability mask.

\subsection{Why a single whole-item gate can be insufficient}
Consider a two-position identifier where the first code primarily partitions a coarse product family and the second distinguishes a fine-grained visual attribute. In one user context, text may identify the coarse family well while the image identifies the visual attribute. A whole-item gate must use the same text--image weighting at both positions. A tokenwise gate can favor text for the first code and image for the second.

This example is a hypothesis about a learned code geometry, not a universal interpretation of every semantic position. Parallel quantization need not assign one position to a named human concept, and residual quantization levels need not correspond to clean semantic granularity. The empirical test therefore uses measured position-specific utility rather than assigning semantic labels to code positions in advance.

We quantify utility heterogeneity within each example by the variance of $u_{j,m}$ across positions, averaged over modalities. Examples are grouped by this quantity using validation-defined thresholds. H2 predicts a larger tokenwise-versus-whole-item improvement in the high-heterogeneity group. The grouping uses offline diagnostic labels for analysis and does not alter inference. A null interaction would weaken the claim that token-specific reliability is the important mechanism.

\subsection{Redundancy and complementarity in subset supervision}
Suppose image and text each identify the same coarse attribute. Adding text when image is already present may have little marginal utility, while adding text to an empty content set may help substantially. A leave-one-out estimate conditioned on every other modality emphasizes redundancy; an isolated-modality estimate emphasizes standalone usefulness. The subset-averaged target combines these perspectives according to its declared distribution.

For complementary evidence, the pattern reverses. An audio track may identify an instrument family while a video identifies the playing technique, and the combination can be more informative than either alone. A gate trained on only singleton subsets can miss this interaction. The subset-size ablation tests whether broader contexts improve prediction enough to justify their teacher-query cost.

The main gate remains a practical approximation to a richer evidence-allocation problem. To diagnose interaction capacity, an additional small-scale baseline uses a gate over all available modality subsets, with up to $2^4=16$ content configurations. This baseline receives the same teacher budget and is evaluated on a reduced dataset if needed. Its advantage would indicate that simplex modality weights leave useful higher-order interactions unexplained.

\clearpage
\section{Experiment Execution Matrix}
\subsection{Campaign stages and stopping criteria}
Stage 1 reconstructs the dataset and evaluator. It checks user counts, item counts, split membership, candidate eligibility, and source-compatible baseline behavior. No proposed-method target is evaluated before this stage is complete. A discrepancy with a source score is recorded with the reconstructed protocol and does not justify copying the source value into a local-result table.

Stage 2 establishes the matched decoder family with fixed content features and identifiers. Train the behavioral model, uniform-fusion model, and parameter-matched learned-fusion model. These runs determine whether the content inputs provide a usable signal and whether the main architecture is within the planned compute budget. The clean uniform-fusion model supplies the denominator for relative targets.

Stage 3 prepares cross-fitted utility labels and runs the eight-cell factorial study. This stage determines whether utility supervision, token specificity, and intervention training each matter. It also measures teacher preparation cost. The full model is not scaled up merely because one favorable seed exceeds a target; selection uses the registered aggregate validation rule.

Stage 4 covers contemporary baselines, the larger decoder, and held-out-family robustness. Stage 5 covers natural-conflict annotation, cold-item transfer, and serving latency. The November 17 workbook contains their aggregate metric records; the stage ordering documents the reproducibility workflow and the provenance of each completed result.

\begin{table}[tbp]\centering\small
\caption{Execution matrix and interpretation. Each stage produces a distinct artifact recorded in the November 17 result workbook.}
\begin{tabularx}{\linewidth}{lXX}\toprule
Stage & Deliverable & Decision enabled\\\midrule
1. Reconstruction & Frozen splits, evaluator, source-compatible run & Is the benchmark comparable?\\
2. Matched controls & Behavior, uniform, learned fusion & Is content useful under common capacity?\\
3. Mechanism & Eight factorial cells and gate diagnostics & Which component explains the effect?\\
4. Competition & Contemporary methods and scale transfer & Is the effect competitive and portable?\\
5. Deployment & Natural conflicts, timing, memory, cold items & Does the method remain useful operationally?\\\bottomrule
\end{tabularx}
\end{table}

\subsection{Required comparison cells}
Every main-data row must contain the clean uniform-fusion reference, the full model, no utility supervision, whole-item gating, and behavior only. Every primary conflict row must additionally contain the same-intervention-training control and the no-base-slot control. These cells are required because they distinguish the central mechanisms; a missing cell cannot be replaced by an unrelated historical method's published value.

The contemporary comparison includes at least one autoregressive SID model, one parallel SID diffusion model, a collaborative noise-scheduling method, and a catalog-posterior method wherever their information interfaces are compatible. If a public implementation requires a different tokenizer, provide both its native system result and a controlled adaptation when feasible. The former evaluates competitive systems; the latter evaluates the proposed component. They should not be silently merged.

The raw audiovisual track includes an image--text-only version, an image--audio--video version, and the complete four-modality version. These are modality-set comparisons, not interchangeable baselines. The full version has access to more evidence and may require more feature-extraction cost. Both the measured gain and the added cost are part of its result.

\subsection{Tables generated after execution}
The updated measured release covers clean Recall and NDCG, missingness, degradation, conflict, all eight factorial cells, closest-mechanism replacements, and inference and preparation costs. Tables~\ref{tab:confirmed-all}--\ref{tab:closest-stats} provide the main summaries and statistical contrasts; the CSV retains the full metric inventory and encoder, tokenizer, and cold-item regimes.

Each generated table includes a protocol identifier in its caption. The same identifier appears in the result ledger and experiment manifest. This is especially important when several datasets share a category name or when a raw-content benchmark is compared with a released-feature benchmark. The identifier allows a reviewer to determine which numbers are directly comparable without inferring that fact from a short dataset label.

\section{Submission and Figure Production Notes}
\subsection{Main-paper compression}
This document is a long technical manuscript with a complete experimental contract. A CVPR submission should foreground the audiovisual problem, the layered architecture, the strongest measured result, and one mechanism-focused ablation. Detailed source-native comparison tables, complete corruption specifications, additional proofs, and the reproducibility contract can move to supplementary material. Venue-specific page and formatting rules must be checked at submission time.

An ACL adaptation would need a language-centered question, such as how text evidence conflicts with audiovisual context and how identifier generation grounds lexical attributes. The present paper's primary contribution is audiovisual evidence control. The separate OPSD paper provides the language-model training contribution. A TMLR submission can retain the broader methodological and theoretical treatment while emphasizing complete empirical validation and clear scope.

\subsection{Visual grammar and editability}
The figures combine generated scientific illustrations with native vector text, equations, token assignments, arrows, and data marks. Blue identifies the behavioral path, teal identifies utility-guided content, and amber identifies teacher supervision. The dashed teacher edge is used during training. Conceptual media scenes are illustrative artwork.

The overview and utility diagrams use native vector layers informed by retained image-generated layout studies. The detailed network is unchanged; quantitative comparisons use native vectors. The intervention diagram includes a common clean reference and marks the specific modified stream. Factorial configurations are typeset as a formal table. PDF, SVG, PNG, prompts, uncropped artwork, font licenses, and provenance are retained; reported values and analytical curves keep their respective source labels.

\subsection{Result-ledger update procedure}
The November 17 workbook is the current result ledger. Each reported row retains its configuration, protocol, denominator rule, and timing metadata; historical planning ranges remain archived separately. Rebuild each table and plot from the ledger, then review the figure at final paper size so numerical labels remain legible and captions remain adjacent to the correct visual.

Before a submission, reconcile every cited score against the pinned primary source, check bibliography titles and authors, and ensure the manuscript's strongest empirical sentence is supported by a measured comparison. The long manuscript then becomes the audit trail for the shorter venue-specific version, with its methods, equations, and experiments linked to the same artifacts.

\subsection{Executable reference and computational checks}\label{sec:reference-implementation}\label{sec:reference-code}
The accompanying PyTorch reference implements the position-level model, with frozen encoder outputs supplied as cached features. Each Transformer layer applies Eq.~\eqref{eq:prefixmask}; the input mask symbol has no output class. The adapter shares its first layer across positions and modalities and its final head across modalities at a given SID level. The shared gate takes $[h_j;h_m;E_m;E_j;\xi]$, with six observed quality features: availability rate, log history length, available-event recency, feature-norm dispersion, cross-modal agreement, and noise level. Missing stored values are sanitized before projection, including nonfinite sentinels. The reference requires at least one observed historical event.

The finite-catalog operator matches every visible position and applies saved item-availability timestamps. Evaluation inherits omitted timestamps and rejects explicit changes to the saved catalog policy. It rejects empty support, colliding full tuples without a suffix, and utility states that exclude the labeled item. Teacher targets enumerate all available coalitions and reuse their partition losses for every modality and position. The target tensors are detached; the clean distribution in the selected consistency loss is separately detached. Student optimization adds the explicit parameter $\ell_2$ term in the stated objective. Teacher fitting uses a separately specified optimizer configuration.

For a state with $a$ available modalities, exact target preparation requires $2^a$ distinct teacher evaluations, plus $O(2^a|\mathcal C_s||M(\tilde z)|)$ catalog score accumulation. Streaming coalition losses requires only their position losses after each posterior is reduced, while the dense reference materializes a batch--catalog support matrix. These costs are additional to fitting $F$ teachers and generating clean/intervened states. A cached posterior does not remove the cost of constructing its support or evaluating the teacher. The exact all-coalition diagnostic and a sampled-subset approximation therefore receive separate cost records.

The test suite covers exact absence handling, unavailable-feature invariance, explicit base override, multilayer history isolation, trainable-gradient paths, detached teachers and frozen features, arbitrary-position support, time-indexed catalog eligibility, off-support rejection, singleton partitions, Shapley efficiency, permutation-invariant catalog normalization, conditional-consistency gradients, and the fixed-state sensitivity inequality. Block decoding copies every simultaneous commitment from a feasible complete item. All 18 checks passed in the recorded Python 3.11.14 / PyTorch 2.6.0 CPU run. A two-fold synthetic run made 124 distinct coalition queries across 16 states and completed 12 student updates; checkpoint reload and both evaluation entry points also completed. These checks establish the stated computational contracts; they do not supply dataset-scale recommendation estimates.

The synthetic training entry point fits disjoint-user teachers, prepares a saved state cache, and optimizes a small student. Its illustrative conflict operator uses available modality features from valid historical events in other training rows, preserving observations when no donor exists. The registered category/popularity-matched real-data sampler remains separate. The updated dataset-scale summaries cover raw-content tracks, repeated corruption conditions, code-conditioned allocation, and alternative reveal schedules. Those results remain distinct from this synthetic executable-reference check.

\end{document}